\documentstyle[aaspptwo,twoside]{article}     

\def\begabst{\begin{abstract}}
\def\endabst{\end{abstract}}
\def\begref{\begin{references}}
\def\endref{\end{references}}
\newfont{\mbf}{cmmib10}
\newfont{\srm}{cmr9}
\def\vec#1{{\bf #1}}
\def\bfeta{\hskip -.5pt\hbox{{\mbf\char17}}}

\begin{document}
\thispagestyle{empty}
\pagestyle{myheadings}
\countdef\sec=21\sec=0
\countdef\subsec=22\subsec=0
\def\section#1{\advance\sec by 1\subsec=0\vspace{2mm}
       \centerline{\number\sec.\ {\footnotesize#1}}\vspace{1mm}}
\def\subsection#1{\advance\subsec by 1\vspace{1mm}
       \noindent\number\sec.\number\subsec.\ {\footnotesize#1}\vspace{1mm}}
\def\begabst{\begin{center}\begin{minipage}{15cm}
       \centerline{ABSTRACT}\vspace{1mm}}
\def\endabst{\end{minipage}\end{center}}
\def\keywords#1{{\it Subject headings:} #1}
\twocolumn[
\vspace{-20mm}\vbox to 17mm{To appear in {\it The Astrophysical
Journal\/}, August 20, 1996 issue, Vol. 467}


\title{THE CONFLUENT SYSTEM FORMALISM:\\ II. THE GROWTH HISTORY OF OBJECTS}
\author{Alberto Manrique$^1$\quad and\quad Eduard Salvador-Sol\'e$^2$}
\affil{Departament d'Astronomia i Meteorologia, Universitat de Barcelona,\\
Avda. Diagonal 647, E-08028 Barcelona, Spain\\
$^1$ alberto@faess2.am.ub.es\quad and\quad$^2$ eduard@faess0.am.ub.es}

\begabst

With the present paper we conclude the presentation of a semianalytical model
of hierarchical clustering of bound virialized objects formed by
gravitational instability from a random Gaussian field of density
fluctuations. In paper I, we introduced the basic tool, the so-called
``confluent system'' formalism, able to follow the evolution of bound
virialized objects in the peak model. This was applied to derive the mass
function of objects. In the present paper, we calculate other important
quantities characterizing the growth history of objects. This model is
compared with a similar one obtained by Lacey \& Cole (1993) following the
Press \& Schechter (1974) approach. The interest of the new modeling
presented here is twofold: 1) it is formally better justified as far as peaks
are more reasonable seeds of objects than the indetermined regions in the
Press \& Schechter prescription, and 2) it distinguishes between merger and
accretion enabling us to unambiguously define the formation and destruction
of objects and to estimate growth rates and characteristic times not
available in the previous approach.


\endabst
]

\section{INTRODUCTION}

Observational data on the distant universe are increasingly numerous
and detailed. Comparison with the nearby universe harbors important
information on the formation and evolution of cosmic objects. However, our
capability of extracting accurate cosmological implications from these
observations is severely limited by the lack of an analytical model of
gravitational clustering. This is a fundamental lack only partially replaced
by numerical simulations.

In the most studied scenario of structure formation via gravitational
instability from a primordial random Gaussian field of density fluctuations,
Press \& Schechter (1974; PS) derived a practical analytical estimate for the
number density of bound virialized objects (or more exactly, steady relaxed
haloes) of mass $M$ at any given epoch. This mass function has been shown
to agree with $N$-body simulations (Nolthenius \& White 1987; Efstathiou et
al. 1988; Efstathiou \& Rees 1988; Carlberg \& Couchman 1989; White et al.
1993; Bahcall \& Cen 1993; Lacey \& Cole 1994). But the mass function of
objects does not provide all the information required in many cosmological
problems. The rates at which objects grow and the characteristic times of
this process are also needed (e.g., Toth \& Ostriker 1992; Richstone, Loeb,
\& Turner 1992; Kauffman, White \& Guiderdoni 1993; Lacey \& Cole 1993;
Gonz\'alez-Casado, Mamon, \& Salvador-Sol\'e 1994; Kauffman 1994;
Gonz\'alez-Casado et al. 1996). Richstone, Loeb, \& Turner (1992) used the
time evolution of the PS mass function to estimate the formation rate of
objects of mass $M$ at different epochs. However, this is not a very
accurate estimate since the time derivative of the mass function is equal to
the rate at which objects reach mass $M$ {\it minus the rate at which they
leave this state\/}, both terms having comparable values.

Following the PS original prescription or a better sound version of it using
the excursion set formalism, Bower (1991) and Bond et al. (1991) derived the
mass function of objects of a given mass at a given epoch subject to the
condition of being part of another object with a larger mass at a later time.
This conditional mass function was used by Lacey \& Cole (1993, LCa; see
also Kauffmann \& White 1993) to infer self-consistent estimates of the
instantaneous merger rate and the typical age and survival time of objects.
This clustering model has been shown by Lacey \& Cole 1994 (LCb) to agree
with $N$-body simulations. However, as recognized by these authors, there is
the formal caveat that the PS approach on which it is based is heuristic. In
particular, the seeds of objects are indetermined, possibly even unconnected,
regions. As a consequence, one cannot really count objects with a given
mass, but just calculate the probability that a given point is in an object
of that mass, which is at the base of a slight inconsistency in the
analytical estimate of the formation time for power spectra with index $n>0$
(see LCa).

But there is perhaps a more serious problem with the PS approach. This deals
with one unique process of instantaneous mass increase of objects, their
capture of other more or less massive ones, process which is generically
called merger. Although strictly correct, this viewpoint is at the base
of an important shortcoming. For usual power spectra of the initial density
field, the number density of objects diverges at small masses. Consequently,
objects continuously experience captures, there being no specific event in
this approach marking the beginning or the end of any stable entity. The age
and survival time of an object must then be defined rather {\it arbitrarily}
in terms of some relative mass variation along the series of embedded objects
connecting with it. Moreover, these definitions are {\it ambiguous} since
they do not account for the fact that the same mass increase can be achieved
in a smooth quasi-continuous manner or essentially in one unique capture at
some intermediate time.

This is particularly disturbing since the notions themselves of the formation
and destruction of objects have been introduced with the aim to specify when
the instantaneous mass increase of a system is important enough to
appreciably affect its structural properties. In this sense, the distinction
between ``tiny'' and ``notable'' captures would be very convenient. Kitayama
\& Suto (1996) have recently proposed a modified version of the LCa model in
this spirit. However, the PS approach is not well adapted to such a
distinction which makes the model of Kitayama \& Suto be somewhat
inconsistent (see \S\ 2 for details). But this does not imply, of course,
that such a distinction cannot be self-consistently achieved in any other
approach.

In tiny captures, the internal kinetic and potential energies of the
capturing system are not appreciably modified from their equilibrium values.
So the system remains essentially unaltered despite the slight extra mass
added. In notable captures (possibly the same preceding events but considered
from the viewpoint of the partner object), there is on the contrary a marked
departure in the internal kinetic and/or potential energies of the capturing
system. This causes an important rearrangement of the system as it relaxes
approaching the virial relation. We can therefore naturally say that, in the
former case, the final and initial relaxed systems are the same object (or
equivalently, that the object survives to the capture) while, in the latter,
they constitute two distinct objects. In other words, we can define accretion
as those tiny captures contributing to the (quasi)continuous mass increase of
objects, and mergers as those notable ones destroying and giving them rise.
Objects are then well-defined as those stable entities surviving, while
smoothly accreting mass, between two mergers marking their formation and
destruction, and the age and survival time of objects would have an obvious
{\it unambiguous\/} meaning.

The problem with this viewpoint is, of course, where to place the division
between tiny and notable captures. Establishing whether there is a
significant rearrangement of the system as a consequence of a capture is
equivalent to establishing whether the capture causes to it a significant
departure from steadiness. So the difficulty in making that division is
similar and, in fact, tightly connected with that encountered in making the
distinction between relaxed and non-relaxed objects (or in identifying the
exact extent of the former ones): deciding whether a region is steady or not.
As far as objects are continuously capturing other more or less massive ones
strict steadiness is never achieved. Consequently, making {\it in practice\/}
any of these decisions requires choosing some confidence level for assessing
the significance of the departure from steadiness, which introduces some
arbitrariness. However, the advantage of working with {\it theoretical
models\/} is that they do not deal with the full complexity of the real
universe but with some suitable idealization of it. For example, the modeling
of gravitational clustering presumes it is possible to tell, in some
unspecified manner, between relaxed and non-relaxed systems. So why could it
not also presume it is possible to tell, in a similar unspecified manner,
between accretion and merger in the sense above?

As shown in paper I (Manrique \& Salvador-Sol\'e 1995; see also
Salvador-Sol\'e \& Manrique 1995), making these two sole assumptions in the
peak model framework leads to one unique, well justified, self-consistent,
clustering model. The basic tool is a new formalism, the so-called confluent
system of peak trajectories, able to follow the filtering evolution of peaks
tracing the clustering evolution of objects (Salvador-Sol\'e \& Manrique
1994). This formalism was applied, in paper I, to derive the mass function of
objects. In the present paper, we apply it to calculate their growth rates
and times. This clustering model is similar to that developed by LCa in the
sense that these quantities are derived from the statistics of the continuous
random density field encountered at a fixed arbitrary epoch after
recombination when fluctuations are still linear (with the growing factor
only a function of time at all interesting scales) and Gaussian distributed,
and the dynamics of dissipationless collapse are approximated by the
spherical model. However, it is physically better motivated since the assumed
seeds of objects are peaks instead of the indetermined regions in the PS
approach. In addition, it allows us to {\it unambiguously\/} define the
formation and destruction of objects and calculate their characteristic
growth rates and times. It is important to remark that there is no freedom in
the model; there are only one or two parameters (depending on the power
spectrum) governing the dynamics of collapse which can be adjusted by fitting
the real mass function of objects. In particular, there is {\it no
arbitrariness\/} associated with the distinction between merger and
accretion.

But what about the connection between this idealized model and real
(simulated or observational) data? As mentioned, the practical distinctions
between relaxed and non-relaxed and between tiny and notable captures require
the choice of two, likely connected, confidence levels. Now, if the model is
good enough it will agree with empirical data {\it for some specific
values\/} or some narrow ranges of values of these confidence levels or any
equivalent pair of parameters, such as the total mean overdensity of the
system, $\langle\rho\rangle/\rho$ with $\rho$ the mean cosmic density, and
the relative mass increase of the system $\Delta M/M$ (both in principle
functions of $M$ and $t$), for the distinctions between relaxed and
non-relaxed objects and between merger and accretion, respectively. These
would be the unspecified manners of carrying out those distinctions
implicitly assumed in the model. This is the scheme followed by the PS
approach for dealing with the non-trivial problem of defining and identifying
relaxed objects. The model here proposed is intended to similarly deal with
the problem of defining and identifying the formation and destruction of
relaxed objects.

The predicted mass function of relaxed objects agrees with the PS one
recovering the results of $N$-body simulations for relaxed haloes identified
as those regions with overdensity $\langle \rho\rangle/\rho$ greater than a
universal threshold (for $\Omega=1$) equal to $178$. Cole \& Lacey (1996)
have shown, indeed, that this corresponds to the maximum extent of regions in
steady state at some reasonable (unspecified) confidence level. The validity
of our model concerning the predicted growth rates and times is harder to
check because no practical distinction has ever been attempted to implement
in numerical simulations between accretion and merger. Nonetheless, a first
hint can be provided by the comparison of our results with those obtained
from the LCa model also checked to agree with numerical simulations.

In \S\ 2 we remind the basic lines of the confluent system formalism. In \S\
3 we derive the instantaneous formation, destruction, and mass accretion
rates of objects. Their typical ages and survival times are calculated in \S\
4. Our results are discussed and summarized in \S\ 5. The notation used in
this paper is the same as in paper I. Since many calculations are based on
results obtained by Bardeen et al. (1986; BBKS) we have kept as close as
possible to the notation introduced by these authors. The main difference
comes from the fact that we deal with three different kinds of densities,
while BBKS only dealt with one. Especial caution must be made in not mixing
them up. Firstly, there are the normal and conditional density functions of
peaks {\it at a fixed filtering scale\/} $R$ per infinitesimal ranges of
$\nu$, i.e., the density contrast $\delta$ scaled to the rms value
$\sigma_0(R)$, and other possible variables. These were already defined in
BBKS and are denoted by a caligraphic capital n just as in that paper. A
minor difference with the notation used in BBKS is that we specify the fixed
value of the filtering scale $R$ as one parameter. For example, we write
${\cal N}(\nu,R) \,d\nu$ instead of ${\cal N}(\nu)\,d\nu$. Second, there are
the normal and conditional density functions of peaks {\it at a fixed density
contrast\/ $\delta$} per infinitesimal ranges of the filtering scale $R$ and
other possible variables. These density functions, already introduced in
paper I, are denoted by a roman capital n with the fixed value of $\delta$ as
one parameter. We write, for example, $N(R,\delta)\,dR$. Finally, and this is
a novelty, there are also density functions per infinitesimal ranges of both
$R$ and $\delta$ (or the corresponding mass $M$ and time $t$, respectively)
and any other extra variable. These are denoted by a capital n in boldface,
for example, ${\bf N}(R,\delta)\,dR\,d\delta$. These different symbols,
${\cal N}$, $N$, and ${\bf N}$, are usually accompanyed by one superindex
specifying the characteristic property (if any) of the peaks involved and the
subindex $pk$ or no subindex at all depending on whether all peaks with that
characteristic property are included or just those of them tracing bound
virialized objects (see \S\ 2), respectively. Hereafter, we assume comoving
units.

\section{THE CONFLUENT SYSTEM FORMALISM}

By assumption, a merger is any {\it discontinuous\/} mass increase along the
temporal series of relaxed systems subtending at each step the preceding one.
(In practice, there must be ``some significant discontinuity in mass'', or
equivalently ``some significant structural rearrangement'', or still ``some
significant departure from steadiness'' between two consecutive relaxed
systems in the discrete, fine enough, temporal series one can consider.) When
an object merges we say that it is destroyed. Note that if an object is
captured by any more massive it automatically merges regardless of whether or
not this more massive object survives to the event. If it does not we say
that the final system subtending the mass of the merging objects is a new
object that has just formed.

In contrast, accretion is any {\it continuous\/} (we assume it also
derivable) mass increase along the temporal series of relaxed systems
subtending at each step the preceding one. (In practice, if there is ``no
significant discontinuity in mass'', or ``no significant rearrangement'', or
still ``no significant departure from steadiness'' along the series of
discrete fine steps considered, one can regard the mass increase as achieved
in a continuous and derivable manner.) We say that an object survives as long
as it evolves by accretion. When an object evolving by accretion is captured
by any more massive object it merges, hence it is destroyed. However, an
object evolving by accretion can capture less massive objects without being
destroyed. In fact, its (continuous) mass increase is made at the expend of
very tiny objects. Only if they capture smaller although notable objects
(yielding a significant discontinuity in its mass increase) they also merge
and are destroyed. This is the kind of mergers mentioned above giving rise to
the formation of new objects.

Notice that these definitions of accretion and mer- ger refer to the object
whose evolution is being followed, not to the event itself. This latter can
be regarded either as merger or as accretion depending on the object under
consideration. The inconsistency of the model proposed by Kitayama \& Suto
(1996) arises from not accounting for this fact. These authors take the
model of LCa and make the extra assumption that any object which results from
another one increasing its mass by less than a factor 2 can be ``identified''
to it. This kind of capture is therefore regarded as accretion, while that in
which the mass increases by more than a factor 2 is regarded as a true
merger. The instantaneous formation/destruction rate of objects of mass $M$
is then defined as the intantaneous total merger rate for final/initial
objects of that mass {\it including only\/} those captures of objects
less/more massive than half/two-times the mass $M$. Now, according to the
definitions adopted for accretion and merger, all these captured objects
merge indeed in the event, while those more/less massive than half/two-times
the mass $M$ only accrete. However, not all those mergers cause the formation
of new objects of mass $M$ as presumed; a large fraction necessarily
contributes to the mass increase {\it by accretion\/} of those objects
identified to the final ones with mass $M$.

Coming back to our model, the peak model ansatz states that there is a
correspondence between peaks of fixed linear overdensity in the filtered
density field at some arbitrary epoch $t_i$ and relaxed objects at the time
$t$. The overdensity $\delta_c$ is assumed to be a decreasing function
$\delta_c(t)$ of the collapse time, and the filtering scale $R$ an increasing
function $R(M)$ of the mass of the resulting objects. Given the assumed
distinction above between accretion and merger for real relaxed objects, that
correspondence automatically leads to the natural identification and
distinction between each other of accretion and merger events {\it in the
filtering process}. A peak on scale $R+\Delta R$, with $\Delta R$ positive
and arbitrarily small, is the result of the evolution by accretion of a peak
on scale $R$ provided only that the volume (mass) subtended by the latter is
essentially embedded within the one subtended by the former. (Strictly, we
should talk about volumes subtended by ``collapsing clouds associated with
peaks'', i.e., the regions surrounding each peak which enclose a total mass
equal to that of the corresponding final object at $t$. But, for simplicity,
we will hereafter say the volumes subtended by peaks. Likewise, we will say
``nested peaks'' instead of ``peaks with nested associated collapsing
clouds''.) Whenever the identification between couples of peaks on contiguous
scales is not possible, that is, whenever there is a discrete jump in scale
between two consecutive embedded peaks, we are in the presence of a merger.
It is important to mention that the filtering of the density field must be
carried out with a Gaussian window for the density contrast $\delta$ of peaks
to diminish with increasing filtering scale $R$ as required by consistency
with the growth in time of the mass of objects. Hereafter we adopt this
particular window.

{}From this identification among peaks on different scales we have that when
an object evolves by accretion its associated evolving peak traces a
continuous and derivable trajectory $\delta(R)$ in the $\delta$ vs. $R$
diagram. (Note that despite the fact that the density field is assumed
continuous and infinitely derivable, peak trajectories are only required to
be one time derivable as the assumed mass increase in time of accreting
objects.) In contrast, when an object merges the evolving peak tracing it
becomes nested on a larger scale peak with identical $\delta$, which yields a
discrete horizontal jump in scale of the associated peak trajectory in the
$\delta$ vs. $R$ diagram. Therefore, to compute the density of objects at $t$
in an infinitesimal range of masses $dM$ we must calculate the density of
{\it non-nested\/} peaks with fixed value of $\delta$ appropriate to $t$ on
scales in the infinitesimal range $dR$ corresponding to $dM$. This density,
$N(R,\delta)\,dR$, can be obtained from the density of peaks satisfying
identical constraints although disregarding whether they are nested or not,
$N_{pk}(R,\delta)\,dR$, and the density of these same peaks subject to the
condition of being located in a background with the same density contrast
$\delta$ on a different filtering scale $R'$, $N_{pk}(R,\delta|R',\delta)
\,dR$. Indeed, these three quantities are related through
\begin{displaymath}
N(R,\delta)=N_{pk}(R,\delta)~~~~~~~~~~~~~~~~~~~~~~~~~~~~~~~~~~~~~~~~~~~~~
\end{displaymath}
\begin{equation}
~~~~~-\int_R^\infty
dR'\,{M(R')\over\rho}\,N(R',\delta)\,N_{pk}(R,\delta|R',\delta),
\end{equation}
with $M(R)$ the inverse relation to $R(M)$. Notice that this relation is
approximated for it assumes the whole cloud associated with the peak on scale
$R'$ with uniform smoothed density contrast equal to its central value.

Equation (1) is a Volterra type integral equation of the second
kind for the unknown function $N(R,\delta)$. This can be obtained by means of
the standard iterative process from the known functions $N_{pk}(R,\delta)$
and $N_{pk}(R,\delta|R',\delta)$, respectively equal to ${\cal
N}_{pk}(\nu,R)$ $\langle x\rangle\,\sigma_2\,R/\sigma_0$ and ${\cal
N}_{pk}(\nu,R|\nu',R')\,\widetilde {\langle x\rangle}\,\sigma_2\,R/\sigma_0$
in terms of the normal and conditional density functions of peaks at a fixed
scale $R$ per infinitesimal range of $\nu=\delta/\sigma_0$, calculated by
BBKS. In this latter expressions $\sigma_i$ are the $i$-th order spectral
moments, which only depend on $R$, and $\langle x\rangle$ and $\widetilde
{\langle x\rangle}$ are some averages of $x\equiv
-(\sigma_2\,R)^{-1}\partial_R\delta$, which depend on $R$ and $\delta$ as
well as on $R'$ in the latter case, defined in paper I. The
mass function of objects at $t$ is then simply
\begin{equation}
N(M,t)=N(R,\delta_c)\,\,{dR\over dM},
\end{equation}
with the dependence on $M$ and $t$ on the right hand side given by the
functions $R(M)$ and $\delta_c(t)$. As shown in paper I, very general
consistency arguments allow one to determine the shape of these two functions
as
\begin{equation}
R(M)={1 \over \sqrt{2\pi}\,q}\,\biggl({M \over \rho}\biggr)^{1/3},
\end{equation}
\begin{equation}
\delta_c(t)=\delta_{c0}\,{a(t_i)\over a(t)}=\delta_{c0}
\biggl({t_i\over t}\biggr)^ {2/3}
\end{equation}
with $a$ the cosmic scale factor and $q$ and $\delta_{c0}$ two constants
governing the exact dynamics of collapse. For CDM and the $n=-2$ power-law
spectra, a good fit is obtained to the PS mass function for $q$ equal to
$\sim1.45$ in both cases (in the power law case, the larger $n$, the smaller
$q$) and $\delta_{c0}$ equal to $\sim 6.4$ and 8.4, respectively. (As
mentioned in paper I, there is no degeneracy in the present formalism between
these two parameters. It is therefore not surprising that a value of $q$
appreciably different from unity is coupled to a value of $\delta_{c0}$ so
different from the standard value of 1.686.) Strictly, equations (3) and (4)
are only valid for the scale-free case, i.e., an Einstein-de Sitter universe
($\Omega=1$, $\Lambda=0$) and a power-law spectrum. However, as shown in
paper I, they are good effective relations for other power spectra such as
the CDM one. Furthermore, following the same strategy as in LCa, it should be
possible to extend the applicability of the model to the cases $\Omega\ne 1$
and/or $\Lambda\ne 0$. We want to emphasize that the values of parameters $q$
and $\delta_{c0}$ quoted above, and used throughout the present paper to
illustrate the practical implementation of the model, correspond to those
giving acceptable fits to the original PS mass function (for the standard top
hat window and critical threshold density) for the same cosmogonies.
This fitting tends to privilege the small mass end while the model will
be finally applied rather to massive objects. Thus, finer values of these
parameters should be inferred by directly fitting $N$-body simulations in the
relevant scales.

Given the characterization of accretion and merger, it is clear that the
density of objects being destroyed (because merging) in a given interval of
time is given by the density of non-nested peaks which become nested along
the corresponding decrement in $\delta$. The density of forming objects is
not so simple to obtain. We first need to characterize those filtering events
which contribute with the {\it appearance\/} of new peaks. As just mentioned,
when an object merges the non-nested peak tracing its evolution in the
$\delta$ vs. $R$ diagram experiences a discrete jump in the scale. But this
does not mean, of course, that every non-nested peak partaking of the same
event necessarily experiences this kind of jump. The largest scale peak will
the most often just keep on evolving in a continuous manner. (This reflects
the well-known fact, in gravitational clustering, that a merger from the
viewpoint of one given object can be seen as accretion for the most massive
partner.) But it will sometimes happen that the largest scale peak also
experiences a discrete jump. Then, it will not be possible to identify the
final non-nested peak with any of its ancestors. This appearance of a new
non-nested peak therefore traces the formation of a new bound object. Hence,
the density of objects forming (because of some mergers) in an interval of
time should be given by the density of peaks appearing in the sense above
along the corresponding decrement in $\delta$.

Actually, things are a little more complicated than this. The continuous
trajectory attached to an accreting non-nested peak can be suddenly truncated
(it is not possible to identify the peak at the current $\delta$ with any
peak on an infinitesimally larger scale) without becoming nested. In
addition, peaks can not only become nested into larger scale ones but they
can also sporadically leave their hosts. According to the correspondence
between relaxed objects and non-nested peaks, the preceding events can only
be interpreted as tracing a relaxed object breaking up in small pieces and a
relaxed object spallating off another object, respectively. Thus, we are
concerned with 5 different filtering processes: accretion (continuous
non-nested peak trajectories), aggregation (continuous non-nested peak
trajectories becoming nested into old or newly appeared continuous non-nested
peak trajectories), fusion (appearance of continuous non-nested peak
trajectories), and the inverse processes of the latter two, namely,
spallation (continuous nested peak trajectories becoming non-nested from
surviving or just disappeared continuous non-nested peak trajectories), and
fission (disappearance of continuous non-nested peak trajectories). Since
spallation and fission are clearly unrealistic processes from the
gravitational viewpoint, the confluent system formalism cannot be used to
follow the clustering of {\it individual\/} objects. However, our purpose is
not to follow the evolution of individual objects, but to own a good {\it
statistical\/} description of the general clustering process.

Therefore, what exactly traces, statistically, the amount of objects that
are destroyed (because merging) or that form (as the result of some mergers)
along some interval of time $\Delta t$ is the {\it net\/} decrease in old
non-nested peak trajectories (given by the amount of aggregations minus
spallations) and the {\it net\/} increase in new non-nested peak trajectories
(given by the amount of fusions minus fissions) yielded along the
corresponding $-\Delta\delta$, respectively. Hereafter, we will simply refer
to these quantities as the net density of peaks becoming nested and the net
density of non-nested appearing peaks, respectively. To calculate the former
we must compute the density of peaks at $\delta$ which in continuously
evolving (without fissioning) to $\delta-\Delta \delta$ are found to be
nested, minus the density of peaks at $\delta$ which are already nested and
do not fission in the next $-\Delta\delta$. Indeed, this latter term includes
peaks which can end up, at $\delta-\Delta\delta$, being nested or non-nested.
Hence, that difference really gives the density of non-nested peaks at
$\delta$ which have just become nested through aggregations, minus the
density of nested peaks at $\delta$ which have just become non-nested through
spallations, as wanted. (We have required all peaks at $\delta$ not to be
fissionable because aggregation and spallation yield, by definition,
continuous peak trajectories.) Similarly, to calculate the latter net density
above we must compute the density of non-nested peaks at $\delta$ which do
not fission in the next $-\Delta\delta$, minus the density of peaks at
$\delta$ which have continuously evolved (without fissioning) from peaks at
$\delta+\Delta\delta$. Indeed, this latter term includes peaks which at
$\delta$ become fissionable or not. Hence, that difference really gives the
density of non-nested peaks at $\delta$ that have just appeared through
fusions, minus the density of non-nested peaks at $\delta$ that are just
going to disappear (or, in the limit of vanishing $\Delta\delta$, have just
disappeared before reaching $\delta$) through fissions, as wanted. (We have
required all the peaks at $\delta$ to be non-nested because fusion and
fission only concern, by definition, peaks which are non-nested in appearing
or disappearing.) Consequently, what we only need for the model to be {\it
statistically} acceptable is those two net quantities tracing the densities
of objects merging and forming along some interval of time be positive.

Let us finally note that mass is usually not conserved in individual
filtering mergers (more exactly, in aggregations or spallations). Indeed, the
mass of an aggregated (spalled) peak does not yield any instantaneous finite
change in the mass of its partner. Only in the case of fusions (fissions)
will the mass of the final (initial) peak be close to that of all aggregating
(spallating) peaks. In any event, the total mass of peaks partaking of the
filtering processes tracing the mergers of objects is certainly conserved
{\it statistically}, as needed. Indeed, the correct normalization at every
$\delta$ of the scale function of non-nested peaks (see paper I) and the fact
that the total available mass in any representative (large enough) comoving
volume is fixed guarantee that the the mass shared by peaks in that volume is
preserved under any instantaneous rearrangement through aggregations and
spallations and the fusions and fissions they experience.

\section{GROWTH RATES}

As just mentioned, the density of objects that are destroyed (because
merging) in the interval $dt$ is traced by the net density of peaks becoming
nested in the corresponding range $-d\delta$. To be more exact, we are
interested in calculating the net density of peaks with $\delta_c$ on scales
between $R$ and $R+dR$ which become nested into (i.e., aggregate into, minus spallate off)
non-nested peaks with $\delta_c-d\delta$ on scales between $R'$ and $R'+dR'$,
${\bf N}^d(R\rightarrow R',\delta_c)\,dR\,dR' \,d\delta$. This net density is
derived in Appendix B. Now, by dividing it by $N(R,\delta_c)\,dR$ we obtain
the net conditional probability that a non-nested peak with $\delta$ on scale
$R$ becomes nested into a non-nested peak with $\delta_c-d\delta$ on scales
between $R'$ and $R'+dR'$. And from this conditional probability we can
readily infer the instantaneous destruction (or true merger) rate at $t$ for
objects of mass $M$ per specific infinitesimal range of mass $M'$ ($M<M'$) of
the resulting objects,
\begin{equation}
r^{d}(M\rightarrow M',t)={{\bf N}^d(R\rightarrow
R',\delta_c)\over N(R,\delta_c)}\,\,{dR'\over dM'}\,\biggl|{d\delta_c\over
dt}\biggr|,
\end{equation}
with $R$, $R'$, and $\delta_c$ on the right hand side written in terms of
$M$, $M'$, and $t$, respectively, through equations (3) and (4). (Note that
this destruction rate involves any possible merger that an object can
experience with other objects, disregarding whether or not this yields the
formation of a new object.)

\begin{figure*}[hbtp]
\centering
\centerline{\epsfxsize= 12cm\epsfysize= 8cm\epsfbox[50 100 550 550]
{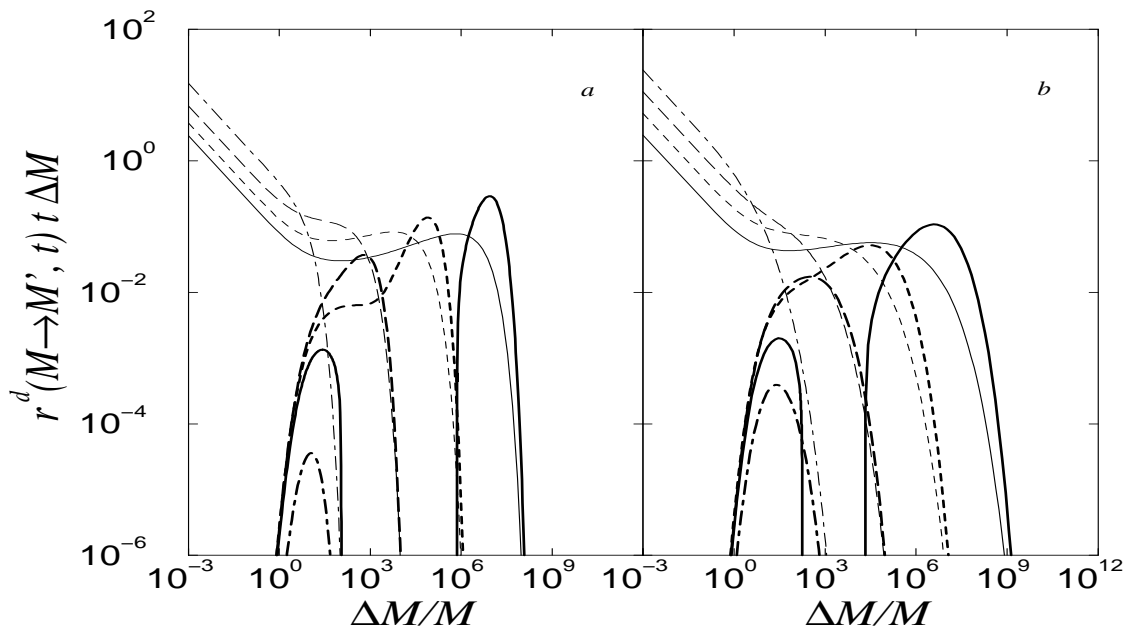}} \caption{\baselineskip=3pt\srm
Instantaneous destruction (or true merger) rate at the present time for
an object with mass $M/M_\odot$ equal to $10^{8}$ (solid lines),
$10^{10}$ (dashed lines), $10^{12}$ (long-dashed lines), and $10^{14}$
(dotted-dashed lines) as a function of the mass $\Delta M$ of the
merger partner for the CDM (a) and $n=-2$ power-law (b) power spectra,
both normalized to $\sigma(8\,h^{-1} \,{\srm Mpc})=0.67$ in an
Einstein-de Sitter universe with $h=0.5$. In thin lines the merger
rates drawn from the LCa model.\label{fig1}\normalsize}
\end{figure*}

This destruction (or true merger) rate is plotted in Figure 1 for different
masses $M$ of the initial object and $t$ equal to the present time. As can be
seen, it shows a very different behavior from the merger rate obtained by
LCa: the former vanishes while LCa's diverges at small $\Delta M/M$. This
marked difference simply reflects the different viewpoints adopted in the two
models. In the PS approach, any mass increase of objects in any given
interval of time is necessarily due to (generic) mergers, while in ours it
can be due to mergers or accretion. So both merger rates cannot be identical.
For small intervals of time, large mass increments are predominantly caused
by mergers rather than accretion. Therefore both models should yield similar
merger rates at very large $\Delta M/M$ as found. Conversely, very small
mass increments should tend to become accretion. In fact, we would even
expect they all become accretion for $\Delta M/M$ below any effective
threshold marking the border line between tiny and notable captures. So it is
well understood that our merger rate vanishes at very small $\Delta M/M$
while LCa's diverges there because of the divergent number of infinitesimal
mass captures. (Note that what LCb compared with $N$-body simulations is not
their merger rate, but the conditional probability that given an object of
fixed mass at some initial epoch it is incorporated into a larger mass object
at a later epoch. Since this quantity does not depend on the particular
definition adopted for merger and accretion, the good agreement found by
these authors between theory and simulations does not favor LCa's viewpoint
as compared to ours.)

Our merger rate therefore shows the good expected behavior at both mass
ends. Moreover, the sharp cut-off shown at small $\Delta M/M$ points to
a division between accretion and merger roughly consistent with a fixed
effective threshold in $\Delta M/M$ (at least for given $M$ and $t$).
However, at relatively large values of $\Delta M/M$, our merger rate is
significantly lower than LCa's which is hard to interpret in terms of our
distinction between accretion and merger. In fact, this effect becomes
increasingly marked as we diminish $M$: a central dip develops finally
reaching negative values. So not only does the correct behavior of the model
seem to require large $M$ but there is a strict lower bound in $M$ for its
general physical consistency. (Negative merger rates imply more spallations
than aggregations which is unrealistic in hierachical clustering.) The value
of this lower bound in $M$ is not dramatic by itself. At present, it is of
the order of the mass of small dwarf galaxies ($\sim 2.5\times 10^9\ M_\odot$
and $\sim 2.5\times 10^8\ M_\odot$, for the CDM and $n=-2$ power-law spectra
here analized, respectively) and it increases with increasing time so that it
encompasses all relevant cosmological scales ($\sim 1.8\times 10^6\ M_\odot$
and $\sim 1.1\times 10^6\ M_\odot$) at a redshift of just $1.25$. What is
disturbing is the own existence of this bound and the related unsatisfactory
behavior of the merger rate at moderate $\Delta M/M$. These shortcomings do
not seem to be inherent to the basic assumption of the model, that is, the
possibility to effectively distinguish between accretion and merger. They
could be partly due to the fact that the peak model is only a good
approximation for the gravitational growth of massive objects. (At any given
$\delta_c$, the larger the scale, the higher the peak amplitude as compared
to the typical fluctuation, $\sigma_0(R)$. And since the higher the peak, the
more spherical its shape and the more negligible the shear caused by the
surrounding matter, the better its gravitational collapse is described by the
spherical approximation.) But what the most likely causes them is the
approximation used in equation (1). Indeed, some preliminary calculations
seem to indicate that a more accurate approximation for the nesting
correction should notably improve the performances of the model. In the
following, we will not insist anymore on the limited domain of consistency of
the present model (in its current version) and focus on the analysis of its
predictions for large enough $M$.

\begin{figure*}[hbtp]
\centering
\centerline{\epsfxsize=12cm\epsfysize= 8cm\epsfbox[50 100 550 550]
{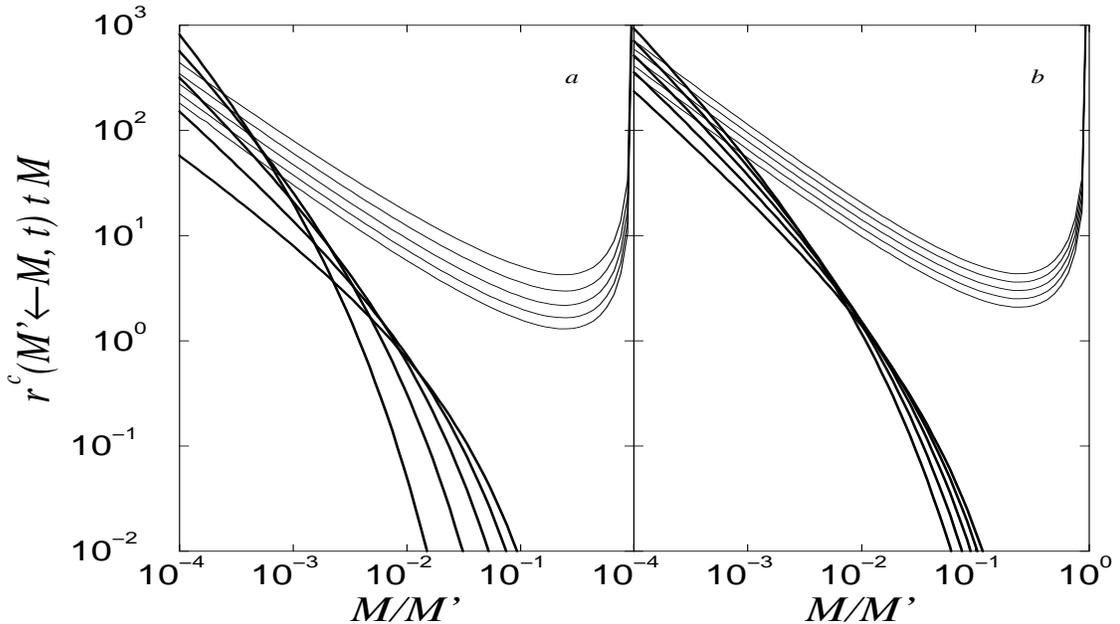}} \caption{\baselineskip=3pt\srm
Instantaneous capture (or accretion+ merger) rate at the present time for a
final object with mass $M'/M_\odot$ equal to $10^{14}$ (the lowest curve on
the left), $3\times 10^{13}$, $9\times 10^{14}$, $2.7\times 10^{15}$, and
$8.1\times 10^{15}$ (the highest curve on the left) as a function of the
mass $M$ of the captured objects for the same cosmogonies (a) and (b) as in
Fig. 1. In thin lines the capture rates inferred from the LCa merger rates.  
\label{fig2}\normalsize}
\end{figure*}

{}From the confluent system formalism making the natural distinction between
accretion and merger we can also calculate the transition rate {\it including
both kinds of captures}, mergers and accretion. Equation (5) gives the rate
at which objects of mass $M$ merge and are destroyed giving rise to objects
of mass $M'$. But from the viewpoint of these latter objects the event can be
seen either as accretion, if they can be identified with the initial most
massive partner, or as a true merger, if they cannot. Thus, the instantaneous
accretion+merger, or simply capture, rate for final objects of mass $M'$ per
specific infinitesimal range of the captured mass $M$ ($M<M'$) is
\begin{equation}
r^{c}(M'\leftarrow M,t)={r^d(M\rightarrow M',t)\,N(M,t)\over N(M',t)},
\end{equation}
with $r^d$ given by equation (5). This capture rate is not directly
comparable to the LCa merger rate plotted in Figure 1 since it refers to the
mass $M'$ of the final object instead of to the mass $M$ of the initial
object. But it can be readily compared to the capture rate drawn from the LCa
merger rate through the same relation (6). This comparison is shown in Figure
2. Once again we find a markedly different behavior between the solution of
the two models. This difference is also due to the distinct viewpoint adopted
in the two models. The capture rate obtained from our model making the
distinction between accretion and merger only includes, in the case of
accretion, those objects which are accreted, obviously not the accreting ones
identified to the final objects of mass $M'$, and hence, not considered as
captured by them. In contrast, all objects are counted in the LCa capture
rate (in particular, those seen as accreted as well as those seen as
accreting from our point of view) because there is no identification in the
LCa model between the final and any of the initial objects partaking of any
capture. This clearly explains the difference between the two capture rates
at captured masses $M$ close to $M'$. On the other hand, the two models
should show a similar behavior at small $M$, in full in agreement with what
is found.

The previous destruction and capture rates are per {\it specific\/}
infinitesimal range of mass of the final or captured objects, respectively.
To derive the respective {\it global} rates we must simply integrate equation
(5) over $M'$
\begin{equation}
r^{d}(M,t)=\int_M^\infty r^d(M\rightarrow M',t)\,dM',
\end{equation}
and equation (6) over $M$
\begin{equation}
r^{c}(M',t)=\int_0^{M'} r^{c}(M'\leftarrow M,t)\,dM.
\end{equation}
Note that these global rates are obviously positive down to much smaller
masses $M$ or $M'$ than the respective specific ones for the whole range
of the extra variable. It is worthwhile mentioning that these global merger
and capture rates cannot be compared with the respective quantities in the
LCa model because these cannot be calculated in that model (the corresponding
integrals diverge). This is an important drawbak of the PS approach related
to the fact no distinction is made between merger and accretion.

\begin{figure*}[hbtp]
\centering
\centerline{\epsfxsize= 12cm\epsfysize= 8cm\epsfbox[50 100 550 550]
{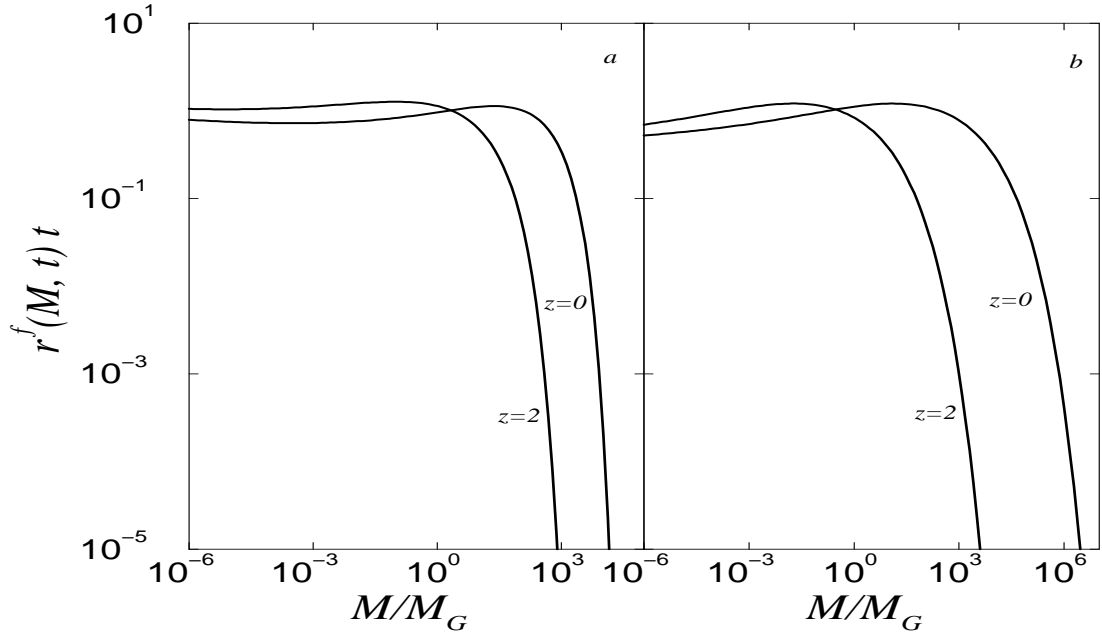}} \caption{\baselineskip=3pt\srm
Instantaneous formation rate at two different epochs as a
function of the mass $M$ (in units of $M_G = 10^{12}\,M_{\odot}$) of objects
for the same cosmogonies (a) and (b) as in Fig. 1.\label{fig3}\normalsize}
\end{figure*}

Let us now turn to the formation rate. The density of objects that form
(as the result of mergers) in the interval of time $dt$ is traced by the net
density of non-nested appearing peaks yielded in the corresponding range
$-d\delta$. More exactly, we are interested in calculating the net density of
non-nested peaks with $\delta_c$ on scales between $R$ and $R+dR$ that just
appear (i.e., fusion minus fission) in the preceding range $-d\delta$,
${\bf N}^f(R,\delta_c)\,dR \,d\delta$. This net density is derived in
Appendix C. Now, by dividing it by $N(R,\delta_c)\,dR$ we are led to the net
conditional probability that a non-nested peak with scale $R$ appears between
$\delta$ and $\delta_c-d\delta$. Therefore, the instantaneous formation rate
at $t$ of objects of mass $M$ is
\begin{equation}
r^{f}(M,t)= {{\bf N}^f(R,\delta_c)\over N(R,\delta_c)} \,
\biggl|{d\delta_c\over dt}\biggr|,
\end{equation}
with $R$ and $\delta_c$ on the right hand side written in terms of $M$ and
$t$, respectively, through equations (3) and (4). This formation rate is
plotted in Figure 3. No similar quantity is provided by the LCa model.

Finally, by multiplying equation (5) by $\Delta M=M'-M$ we obtain
the specific rate at which the mass $M$ of objects is increased at the
time $t$ owing to (true) mergers and, by integrating this latter function
over $M'$, the instantaneous typical mass increase rate owing to mergers for
objects of mass $M$
\begin{equation}
r^{d}_{mass}(M,t)=\int_M^\infty\,\Delta M\,r^{d}(M\rightarrow M',t)\,dM'.
\end{equation}
Likewise, from equation (6) we can obtain the instantaneous typical mass
increase rate for objects of final mass $M'$ owing to accretion and merger,
\begin{equation}
r^{c}_{mass}(M',t)=\int_0^{M'}\,M\,r^{c}(M'\leftarrow M,t)\,dM.
\end{equation}

\begin{figure*}[hbtp]
\centering
\centerline{\epsfxsize= 12cm\epsfysize= 8cm\epsfbox[50 100 550 550]
{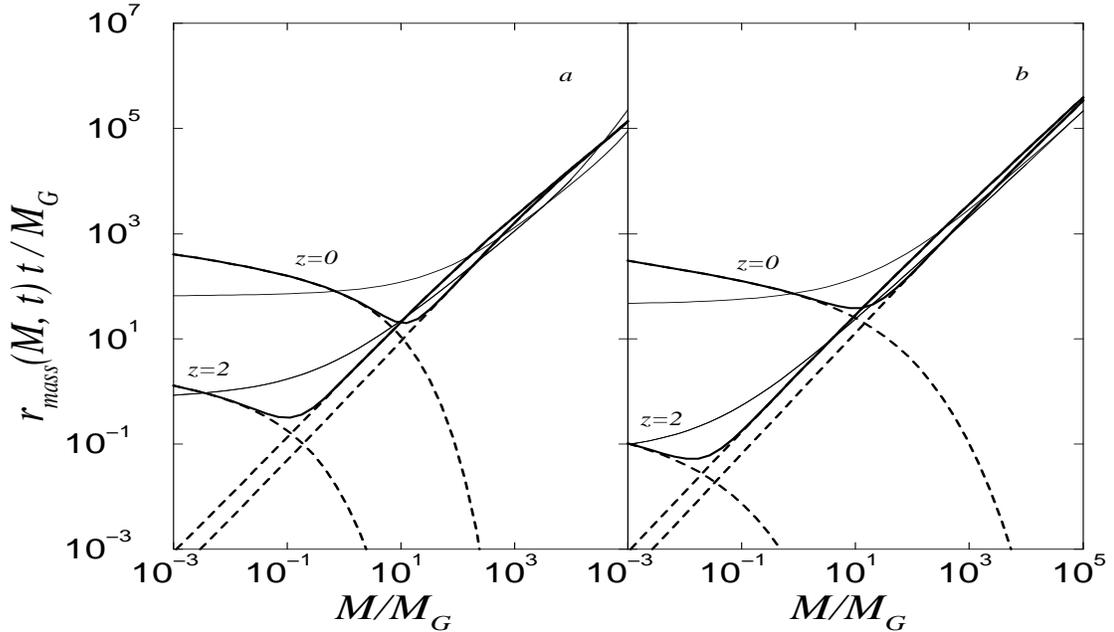}} \caption{\baselineskip=3pt\srm
Instantaneous total mass increase rate for the same epochs and masses
as in Fig. 3. The contributions from mergers and accretion (dominating
at the lower and higher mass ends, respectively) are represented in
dashed lines. In thin lines the total mass increase rates inferred from
LCa merger rates. \label{fig3}\normalsize}
\end{figure*}

As shown in Appendix D, the idea of a {\it transition} rate, by accretion,
similar to those given by equations (5) or (6) is meaningless because
accretion is a continuous instead of a discrete process of mass increase. In
contrast, the {\it mass increase} rate by accretion does make sense. The
instantaneous mass accretion rate for objects of mass $M$ follows from the
instantaneous scale increase rate of the corresponding peaks as they evolve
along continuous and derivable trajectories in the $\delta$ vs. $R$ diagram.
This latter rate depends on the particular value of the scaled Laplacian $x$
of the peak which is being followed. However, since we are interested in the
{\it typical} mass accretion rate for objects of mass $M$ disregarding any
other property we must average over $x$. This is done in Appendix D. The
result is
\begin{equation}
r^a_{mass}(M,t)\approx
{1\over \langle x\rangle\,\sigma_2\,R}\,\,{dM\over dR}
\,\biggl|{d\delta_c\over dt}\biggr|,
\end{equation}
with $R$ and $\delta_c$ on the right hand side written in terms of $M$ and
$t$, respectively, through equations (3) and (4) and $\langle x\rangle$ the
average mentioned after equation (1).

In Figure 4, we plot the total mass increase rate of objects of mass $M$ at
$t$, $r_{mass}(M,t)\equiv r^d_{mass}(M,t)+r^a_{mass}(M,t)$, and compare it
with that resulting from the LCa model, i.e., the integral over $\Delta M$ of
their specific merger rate times the corresponding increase in mass. (Note
that this latter integral does not diverge contrary to those involved in the
calculation of the total merger and total capture rates.) As expected, the
two models show, in this case, similar qualitative behaviors. They only
differ quantitatively in the regime dominated by mergers. This departure is
just due to the differences in the two merger rates at large $\Delta M/M$
(see Fig.1).

\section{GROWTH CHARACTERISTIC TIMES}

{}From the meaning of the global merger rate, equation (7), we have that
the density $N_{sur}(t)\,dM$ of objects surviving (i.e., having not merged
but just accreted) until the time $t$ from a typical population
with masses in the range between $M_0$ and $M_0+dM$ at $t_0<t$ is
given by the solution of the differential equation
\begin{equation}
{d N_{sur}\over dt}=- r^{d}[M(t),t]\,N_{sur}(t),
\end{equation}
with initial condition $N_{sur}(t_0)=N(M_0,t_0)$. In equation (13) and
hereafter, the function $M(t)$ is the typical mass at $t$ of such
accreting objects, approximately given by the solution of the differential
equation
\begin{equation}
{dM\over dt}=r^a_{mass}[M(t),t],
\end{equation}
with initial condition $M(t_0)=M_0$. Indeed, from equation (12) we have that
the typical mass of objects evolving by continuous accretion increases with
time according to equation (14). This expression is only approximated because
the average over $x$ leading to equation (12) presumes all objects with the
same mass, while the mass $M(t)$ is just the so estimated {\it typical\/}
mass of objects at $t$. Hence, the larger the interval of time spent since
$t_0$ when objects have really identical mass the poorer will be the
approximation.

The solution of equation (13) is
\begin{equation}
N_{sur}(t)=N(M_0,t_0)\,{\rm exp}
\Biggl\{-\int_{t_0}^t r^d[M(t'),t']\,dt' \Biggr\}.
\end{equation}
Hence, by defining the typical survival time $t_{sur}$ of objects
with masses between $M_0$ and $M_0+dM$ at $t_0$ as the interval of time
since $t_0$ after which their initial density is reduced (owing to mergers)
by a factor $e$, we are led to the equality $t_{sur}=t_d-t_0$, with the
destruction time $t_d(M_0,t_0)$ given by the solution of the
implicit equation
\begin{equation}
1=\int_{t_0}^{t_d(M_0,t_0)}\,r^d[M(t'),t']\,dt'.
\end{equation}
In addition, the typical mass accreted by those objects until they merge and
disappear is $M[t_d(M_0,t_0)]-M_0$. (Caution: what LCa called survival time
rather corresponds to what here is called destruction time; hereafter we
assume these authors using the same notation as ours.)

In a fully similar manner we can infer the typical age of objects with
masses in the range between $M_0$ and $M_0+dM$ at $t_0$, that is, the typical
interval of time since the last merger giving them rise. The density
$N_{pre}(t)\,dM$ of these objects pre-existing (i.e., having just accreted
matter since then) at a time $t<t_0$ is given by the solution of the
differential equation
\begin{equation}
{d N_{pre}\over dt}= r^{f}[M(t),t]\,N[M(t),t]-r^d[M(t),t]\,N_{pre}(t)
\end{equation}
with initial condition $N_{pre}(t_0)=N(M_0,t_0)$.
The solution of equation (17) is
\begin{displaymath}
N_{pre}(t)=N(M_0,t_0)\,K^{-1}(t,t_0)~~~~~~~~~~~~~~~~~~~~~~~~~~~~
\end{displaymath}
\vspace{-1.72\baselineskip}
\begin{mathletters}
\begin{eqnarray}
&~&\hskip -16mm\times\biggl\{1  -\int_t^{t_0}\hskip-2mm dt'\,
{N[M(t'),t']\over N(M_0,t_0)}\,r^f[M(t'),t']\, K(t',t_0) \biggr\}.\\
&~&\hskip -4mmK(t,t_0)={\rm exp}\biggl\{-\int_t^{t_0}
r^d[M(\xi),\xi]\,d\xi\biggr\} \end{eqnarray}
\end{mathletters}

\begin{figure*}[hbtp]
\centering
\centerline{\epsfxsize= 12cm\epsfysize= 8cm\epsfbox[50 100 550 550]
{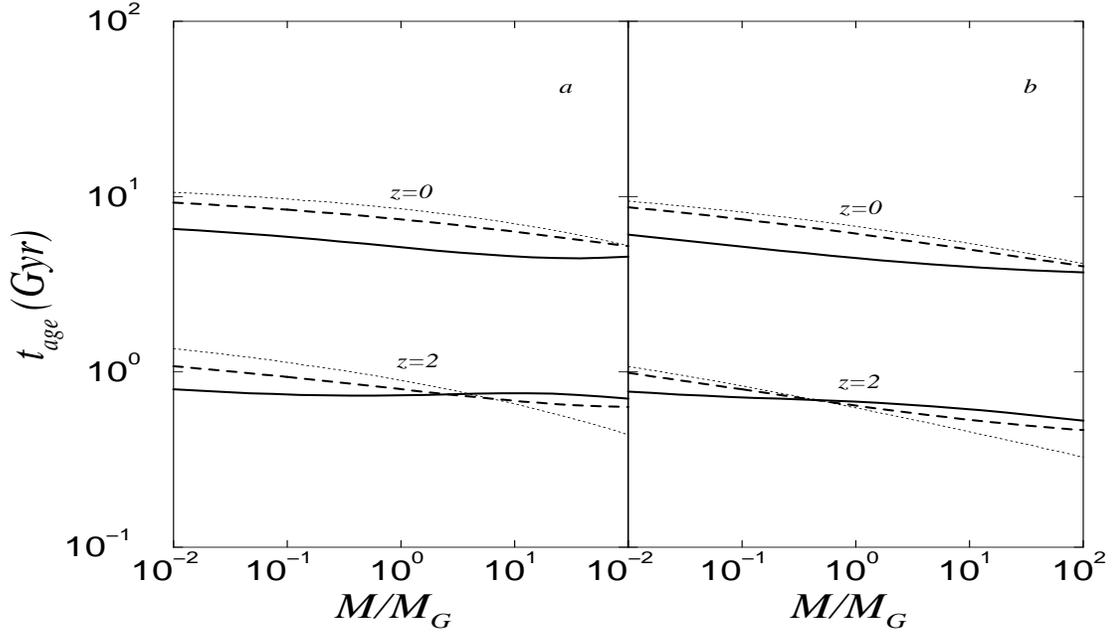}} \caption{\baselineskip=3pt\srm
Typical age (solid lines) and half-mass-accretion time (dashed lines) of
objects of mass $M$ in units of $M_G = 10^{12}\, M_{\odot}$ for the quoted
values of the redshift corresponding to the same cosmogonies (a) and (b) as
in the preceding figures. LCa's median age is plotted for comparison (thin
dotted lines).\label{fig4}\normalsize}
\end{figure*}

Thus, by defining the typical age $t_{age}(M_0,t_0)$ of objects with masses
between $M_0$ and $M_0+dM$ at $t_0$ as the interval of time until $t_0$
before which their density (owing to their progressive formation) was a
factor $e$ smaller, we are led to the equality $t_{age}=t_0-t_f$, with the
formation time $t_f(M_0,t_0)$ given by the solution of the implicit equation
\begin{equation}
1= {N_{pre}[t_f(M_0,t_0)]\over N(M_0,t_0)}\,{\rm e}.
\end{equation}
And the typical mass accreted by these objects since they formed is
$M_0-M[t_f(M_0,t_0)]$.

Finally, from the typical age and surviving time of objects with masses
between $M_0$ and $M_0+dM$ at $t_0$ calculated above, we can readily
calculate their typical lifetime or intermerger period. This is simply
\begin{displaymath}
t_{life}(M_0,t_0)=t_{age}(M_0,t_0)+t_{sur}(M_0,t_0)~~~~~~
\end{displaymath}
\vspace{-1.3\baselineskip}
\begin{equation}
~~~~~~~=t_m(M_0,t_0)-t_f(M_0,t_0).
\end{equation}
And connected with this latter quantity, there is the total mass typically
accreted by those objects during their whole life, given by
$M[t_m(M_0,t_0)]-M[t_f(M_0,t_0)]$.

\begin{figure*}[hbtp]
\centering
\centerline{\epsfxsize= 12cm\epsfysize= 8cm \epsfbox[50 100 550
550]{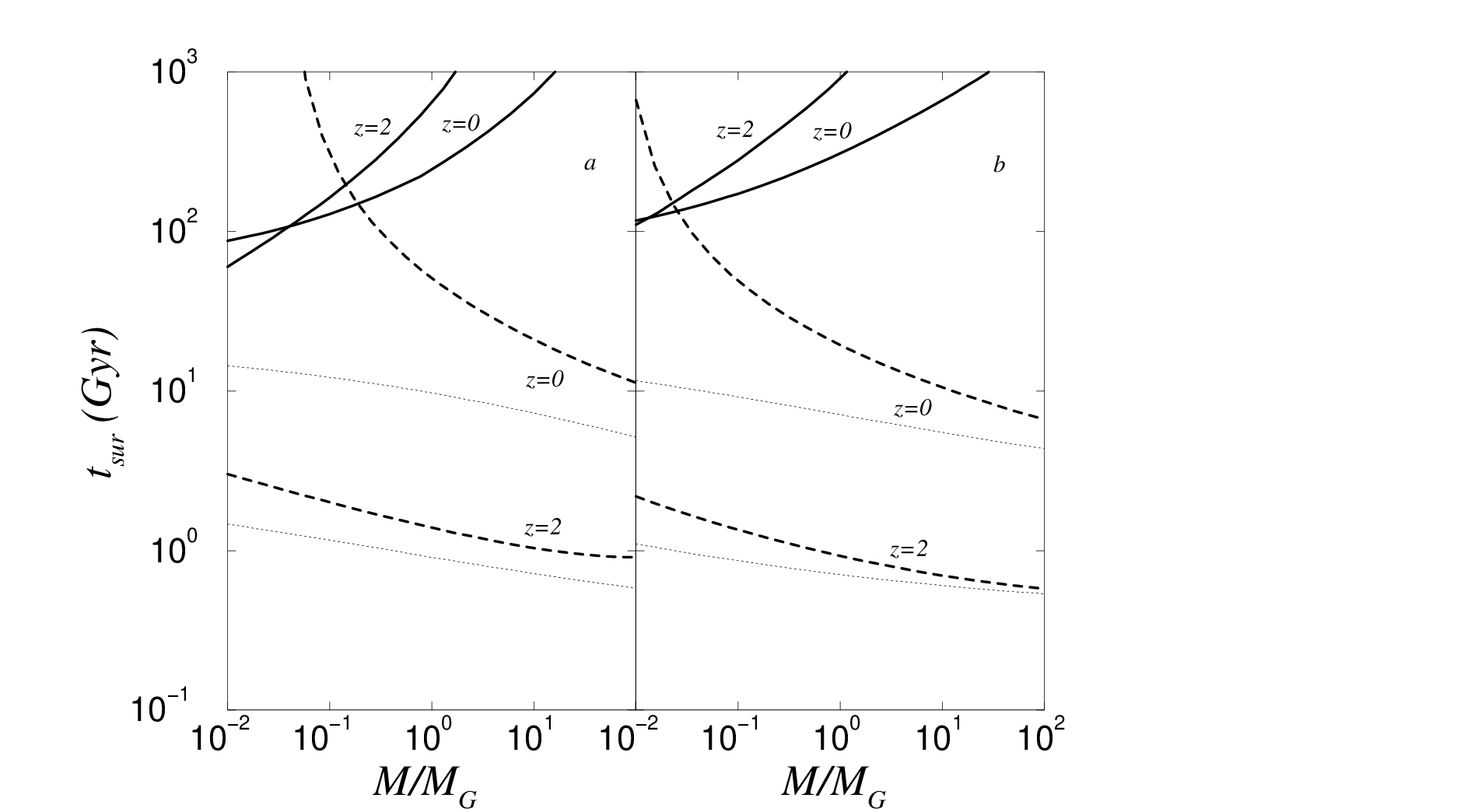}} \caption{\baselineskip=3pt\srm
Typical survival time (solid lines) and double-mass-accretion
time (dashed lines) of objects of mass $M$ in units of $M_G =
10^{12}\,M_{\odot}$ for the quoted values of the redshift correponding
to the same cosmogonies (a) and (b) as in Fig. 4. LCa's median
survival time is plotted for comparison (thin dotted lines).
\label{fig5}\normalsize} \end{figure*}

In Figures 5 and 6 we plot the typical age and survival time, respectively,
of objects at two different epochs as a function of their mass $M$. For
comparison we also plot the corresponding time estimates drawn from LCa
model. As mentioned, our natural definitions of the typical age and survival
time (i.e., the time spent since the previous merger when the object was
formed and the time spent until the next merger when the object will be
destroyed, respectively) are not possible in the PS approach not making the
distinction between merger and accretion. What LCa call the age and survival
time of an object of mass $M$ are the time spent since the mass of some
parent object reached half that mass and the time required by the object to
be found inside an object of double mass, respectively. (LCa give the time
distributions; what are plotted in Figs. 5 and 6 are the corresponding median
values.)

Our age shows the same slightly decreasing trend with increasing $M$ and the
same strongly increasing trend with increasing time as LCa's. The only
difference is that our age is slightly smaller than LCa's. (This trend is
reversed at very large masses, for a given redshift, but this is likely due
to the poor fit to the PS mass function yielded there by our mass function.
Note that a mass of $10^{12}$ M$_\odot$ at $z=2$ is comparable to a present
mass of two or three orders of magnitude higher for the $n=-2$ power-law and
the CDM spectra, respectively.) In contrast, our survival time show a very
different, in fact opposite, behavior from LCa's. Our result that the larger
the mass of the object, the larger its survival time is just what one would
expect in hierarchical clustering distinguishing between accretion and
merger. Indeed, at a fixed time, massive objects are rare and they hardly
merge, while light objects are more frequent and can more easily merge (see
Fig. 3). Moreover, as the time increases objects of a given fixed mass become
more and more frequent, hence, their survival time diminishes as also found.

To better understand these comparative behaviors we also plot, in Figures 5
and 6, two new times, hereafter called half-mass-accretion time and
double-mass-accretion time, defined \`a la LCa but in the context of our
model. These are the interval of time spent since the mass of an object was
typically half its current value and the interval of time required by an
object to typically double its mass, respectively. As these new times refer
to the typical mass evolution of {\it given objects\/} the evolutionary
process involved can only be accretion. So they are given by the function
$M(t)$ solution of equation (14). It is worthwhile noting that these new
times are, in some sense, orthogonal to our age and survival times because
the former only depend on accretion while the latter mostly do on mergers.

{}From Figure 5 we see that objects found at a given redshift typically form
at a lookback time of the order of (just slightly smaller than) that required
for them to accrete half their mass. (Note that our age takes into account
that many similar objects that formed earlier or later than this epoch have
not survived until the redshift considered.) So mergers play an important
role in the past evolution of these objects, but the mass increase
experienced by them from the time they typically formed is close to a factor
two. It is therefore not surprising that the age estimate \`a la LCa
(accounting for both accretion and mergers) is also close to that time.
Notice, however, that the similarity between the half-mass-accretion time and
the age as defined in the present work is fortuitous since both time
estimates are rather orthogonal. So is the similarity between our age and
LCa's.

{}From Figure 6 we see that the time required by very massive objects at
a given epoch to be destroyed is much greater than that required for them to
double their mass by accretion, while the opposite is true for very light
objects. In other words, there is a gradation from frequent light objects
where the mass increase is dominated by mergers to rare massive objects where
it is by accretion (see Fig. 3). It is therefore not surprising that the
double-mass-accretion time (only accounting for accretion) and the LCa
survival time (accounting for accretion and mergers) notably deviate from
each other for very light objects while they yield similar results for very
massive ones. Note that we should not expect a very good agreement there,
either, because the larger the resulting value of the time estimate, the more
marked the effects of: 1) the approximation used (eq. \lbrack 14\rbrack) to
calculate accretion-times and 2) the difference between the two couples of
time estimates; ours reports on ``the time required for objects to change
their typical mass by a given factor'', i.e., they involve averages along $R$
at fixed $\delta$'s, while LCa's reports on ``the typical time required for
objects to change their mass by the same given factor'', i.e., they involve
averages along $\delta$ at fixed $R$'s. In any event, the opposite behavior
shown by our survival time as compared to LCa's is also clarified.

\section{DISCUSSION AND CONCLUSIONS}

The idea that the present clustering model relying on the peak model ansatz
can provide a good description of hierachical clustering is at variance with
the rather extended opinion that peaks are not good seeds of bound objects.
Detailed $N$-body simulations of the gravitational evolution of individual
density maxima seem to show, indeed, that if their initial amplitude is small
they can be disrupted before collapsing (van de Weygaert \& Babul 1993).
While Katz, Quinn, \& Gelb (1993) have found that even high amplitude peaks
seem to be poor tracers of bound objects, individually as well as
statistically. The reason why this would be so is that peaks are not
spherical in general, nor are they isolated. So their non-linear evolution
will markedly deviate from the spherical approximation at the base of the
peak model ansatz. However, any conclusion drawn from the previous $N$-body
simulations dealing with peak statistics is somewhat precipitated because of
the unknown effects of the use of unappropriate filtering windows and/or
$R(M)$ and $\delta_c(t)$ relations, and the lack of any correction for the
cloud-in-cloud effect. In fact, by using the accurate peak-patch formalism,
Bond \& Myers (1996) have found that there is actually a good correspondence
between peaks and high mass objects. This is well understood: the higher the
peak relative to the typical density fluctuation on that scale, the more
spherical, and the less important the shear caused by the surrounding density
fluctuations, particularly if the power spectrum is steep enough
($\sigma_0(M)$ flat enough) to guarantee that nearby density fluctuations on
similar scales are typically negligible. This has also been recently
confirmed by Bernardeau (1994) who has rigurously shown that the evolution of
high amplitude peaks (with $\nu\gtrsim 2$) in a Gaussian random density field
is correctly described by the spherical model provided only the logarithmic
slope of the power spectrum is smaller than $n=-1$.

A fundamental result of paper I was that the self-consistency of the peak
model ansatz, in an extended version presuming some effective distinction
between accretion and merger, completely fixes the filtering window and the
$R(M)$ and $\delta_c(t)$ relations except for a couple of free parameters
governing the dynamics of collapse. After calculating the density of peaks at
a fixed $\delta_c$ and performing the appropriate correction for the
cloud-in-cloud effect we obtained a mass function which is in overall
agreement with $N$-body simulations (since in agreement with the PS mass
function) for appropriate values of those parameters.

In the present paper we have calculated the growth rates and characteristic
times of relaxed objects predicted by the model. As far as we can tell from
the comparison with the LCa model not making the distinction between merger
and accretion, our model yields a reasonably good detailed
description of hierarchical clustering for massive objects. Its
unsatisfactory behavior at small $M$ is likely due to one practical although
rather crude approximation used in paper I to correct for the nesting among
peaks at a fixed density contrast. We are currently working on an improved
version of the model based on a more accurate correction.

Let us finally insist on the great potential of the approach followed here.
The violent relaxation that accompanies the sudden large mass increase of a
relaxed system is expected to leave perdurable imprints in its morphological
properties as compared to the effects of a quasi-continuous one. A detailed
clustering model should therefore make that important distinction. This can
be readily carried out in the extreme cases of very large or very small mass
captures, but in the intermediate regime it is more problematic. Do we have
we the right to assume some definite effective division between tiny and
notable captures? Is it unique and where is it? The assumption that there is,
indeed, a definite division between accretion and merger yields, in the well
motivated framework of the peak theory, one unique, self-consistent,
reasonably good clustering model. Thus, the possibility of carrying out such
a division in a self-consistent manner seems proved. The comparison with
$N$-body simulations should tell us where the effective division between
merger and accretion is, opening in this way the possibility of better
describing the growth of cosmic objects and of understanding their observed
properties.

\acknowledgments

We are grateful to the referee, Cedric Lacey, for his fruitful criticisms.
This work has been supported by the Direcci\'on General de Investigaci\'on
Cient\'\i fica y T\'ecnica under contract PB93-0821-C02-01.

\onecolumn
\eqsecnum
\onecolumn
\markboth{MANRIQUE AND SALVADOR SOL\'E~~~~~~~~~~~}{Growth History of Objects}
\appendix
\centerline{\footnotesize APPENDIX A: PREPARATORY RESULTS}
\setcounter{section}{1}
\noindent{\footnotesize A.1. DISAPPEARING PEAKS}

{}From the Taylor series expansion of the gradient of the density contrast
smoothed on scale $R$, $\bfeta$, around the location $\vec r_p$ of a
neighboring peak we have (to first order in $|\vec r-\vec
r_p|$)
\begin{equation}
\eta_i\approx -\lambda_i\,(r-r_p)_i,
\end{equation}
with $\lambda_i>0$, the eigenvalues of the second order cartesian derivative
tensor $\zeta_{ij}$ changed of sign evaluated at $\vec r_p$ or \vec r. Since
there is at most one peak in the neighborhood of any point the density of
peaks around a point $\vec r$ is
\begin{equation}
\delta^{(3)}(\vec r-\vec r_p)=|\lambda_1 \lambda_2
\lambda_3|\,\delta^{(3)}\bfeta.
\end{equation}
But $\eta_i$ and $\lambda_i$ are random Gaussian variables. So the typical
density of peaks on scale $R$ around an arbitrary point \vec r is given by
the mean
\begin{equation}
\langle\delta^{(3)}(\vec r-\vec r_p)\rangle=\langle|\lambda_1 \lambda_2
\lambda_3|\,\delta^{(3)}\bfeta\rangle
\end{equation}
for the joint probability of the random variables involved, with $\lambda_i$
strictly positive, evaluated at the arbitrary point. This is the scheme
followed by BBKS for obtaining the density ${\cal N}_{pk}(\nu,R)$ $d\nu$ of
peaks with scaled density contrast $\nu$ in an infinitesimal range.

According to the identification criterion among peaks on contiguous scales, a
peak at $\vec r_p$ on scale $R$ will disappear in reaching the scale
$R+\Delta R$, with $\Delta R$ positive and arbitrarily small, provided only
this is the first scale larger than $R$ with no peak in the neighborhood
$|\vec r-\vec r_p| \le O(\Delta R)$ of the former peak. From the Taylor
series expansion of ${\bfeta}$ at the nearest point $\vec r_{p'}$ with
${\bfeta}=0$ on scale $R+\Delta R$ around the peak $\vec r_p$ on scale $R$ we
have
\begin{equation}
(r_{p'}-r_p)_i\approx {\partial_R \eta_i\over \lambda_i}\,\Delta R.
\end{equation}
Thus, on the new scale $R+\Delta R$ there is some peak in the neighborhood
$|\vec r-\vec r_p| \le O(\Delta R)$ of the old peak on scale $R$ provided
only that all $\lambda_i$ are of order unity (this also guarantees that all
$\lambda_i$ are positive at $\vec r_{p'}$ on scale $R+\Delta R$). Strictly,
some $\lambda_i$ could be of order $\Delta R$ or smaller, if the
corresponding $\partial_R \eta_i$ were too. But the probability for this
to happen is negligible as compared to the more general preceding case.
Therefore, for that condition to be broken for the first time at $R+\Delta
R$, some eigenvector $\lambda_i$ must become of order $\Delta R$ or,
equivalently, must vanish at $R\pm \Delta R$.

Consequently, the density of peaks on scale $R$ at an arbitrary point
\vec r disappearing at $R'$ in the neighborhood of $R$, with $R'>R$, is given
by the mean $\langle\delta^{(3)}(\vec r-\vec r_p)\rangle$, with all
eigenvalues $\lambda_i$ positive, as in BBKS, {\it and} the smallest one
satisfying the relation
\begin{equation}
R'-R\approx {\lambda_i\over |\partial_R \lambda_i|}.
\end{equation}
But condition (A5) implies, at the same time, that the smallest eigenvalue
$\lambda_i$ also vanishes in the neighborhood of $\vec r_p$ on scale $R$.
Thus, in neglecting second order terms in equation (A1) the term in that
component $\lambda_i$ must be neglected, too. Accordingly, were we interested
in calculating the density of peaks disappearing at $R'$ in the neighborhood
of $R$, we should take for that component of $\vec r -\vec r_p$, instead of
(A1), the relation
\begin{equation}
\partial_R \eta_i -\partial_R \eta_i (\vec r_p) \approx
-\partial_R \lambda_i\, (r-r_p)_i,
\end{equation}
with $\partial_R \eta_i (\vec r_p)$ evaluated at the peak on scale $R$ and
the remaining variables evaluated at the arbitrary point $\vec r$.

Following these schemes it can be shown (Manrique 1995) that the density of
peaks at $R$ with $\nu$ in an infinitesimal range and all eigenvalues
$\lambda_i$ of zeroth order in $\Delta R$ is equal to the total density
${\cal N}_{pk}(\nu,R)\, d\nu$ calculated by BBKS to zeroth order in $\Delta
R$, while the density of peaks with some $\lambda_i$ of first order in
$\Delta R$ contribute to that total density with first order terms in $\Delta
R$. Therefore, the density of peaks at $\delta$ with $R$ in an infinitesimal
range, $N_{pk}(R,\delta)\,dR$, calculated in paper I from the density ${\cal
N}_{pk}(\nu,R)\, d\nu$ is indistinguishable to leading order in $\Delta R$
(or in $\Delta \delta$ for each given trajectory $\delta(R)$) from the
density of peaks at $\delta$ robust enough for their continuity to be
guaranteed in some neighborhood $\Delta R$ (or $\Delta \delta$).

\vspace{\baselineskip}
\noindent{\footnotesize A.2. DENSITIES OF PEAKS WITH SPECIFIC VALUES OF
RANDOM VARIABLES}

In next Appendixes we will be concerned with densities of peaks with values
of the random Gaussian variables with null mean $v_0\equiv\nu$, $v_1\equiv x=
-(R\sigma_2)^{-1}\partial_R(\sigma_0\nu)$ (this is the same variable $x$
defined in BBKS; see paper I), and
\begin{equation}
v_i\equiv{1\over \sigma_{2i}\,R^{i-1}}\,\,\partial_R
[\sigma_{2(i-1)}\,R^{i-2}\,v_{i-1}]\qquad (i\ge 2)
\end{equation}
on scale $R$ in infinitesimal ranges. All these densities can be
inferred from the density of peaks with $\nu$ and $x$ on scale $R$
in infinitesimal ranges (BBKS)
\begin{equation}
{\cal N}_{pk}(\nu,x,R)\,d\nu\,dx=\,{\exp(-\nu ^2/2) \over (2\pi )^2R_*^3\,
[2\pi\,(1-\gamma^2)]^{1/2}}\,f(x)\, \exp \biggl[-{(x-\gamma\nu)^2
\over 2(1-\gamma^2)}\biggr]\,d\nu\,dx
\end{equation}
with $f(x)$ a function given by BBKS (eq. \lbrack A15\rbrack),
$\gamma\equiv \sigma_1^2/(\sigma_0\sigma_2)$,
and $R_*\equiv \sqrt{3}{\sigma_1/\sigma_2}$.
One must simply apply the recursive relation
\begin{equation}
{\cal
N}_{pk}(v_0,v_1,...,v_n,R)\,dv_0\,dv_1\,...dv_n
 = {\cal
N}_{pk} (v_0,v_1,...,v_{n-1},R)\,dv_0\,dv_1,...dv_{n-1}
{\cal P}(v_n,R|v_0,v_1,...,v_{n-1},R)\,dv_n,
\end{equation}
with ${\cal P}(w,R|v_0,v_1,...,v_{n-1},R) \,dw$ the conditional probability
of finding the value of $w$ in an infinitesimal range given that
$v_0,v_1,...,v_{n-1}$ take some given values,
\begin{mathletters}
\begin{eqnarray}
~~~~~~~~~~{\cal P}(w,R|v_0,v_1,...,v_{n-1},R)\,dw\hskip -5mm&=&\hskip
-5mm{1\over \sqrt{2\pi}\,\sigma_w}
\,\exp \biggl[-{(w -\bar w)^2\over 2\,\sigma^2_w}\biggr]\,dw\\
\bar w={\bf M}_{w\,v}\,{\bf M}_{v\,v}^{-1}\,{\bf V}^{\rm T}&\qquad&
\sigma_w^2=<w^2>-\,{\bf M}_{w\,v}\,{\bf M}_{v\,v}^{-1}\,{\bf M}_{w\,v}^{\rm
T},
\end{eqnarray}
\end{mathletters}
where ${\bf M}_{w\,v}$ and ${\bf M}_{v\,v}$ are the $1\times n$
and the $n\times n$ correlation matrixes of $w$ with $v_i$,
$\langle w\,v_i\rangle$, and $v_i$ with themselves, $\langle
v_i\,v_j\rangle$, respectively, ${\bf
V}$ stands for the $1\times n$ matrix of components $v_i$, and ${\rm T}$
denotes transpose. Note that, although equation (A9) involves, in principle,
the conditional probability for peaks, the conditional probability (A10) is
for simple points. The reason for this is that the condition for a point to
be a peak refers to the values of variables $\eta_i$, $x$, $y$, and $z$
(see BBKS for the definition of $y$ and $z$). While none of the variables
$v_i$ defined above correlates with any of these variables except for $x$.
Consequently, the conditional probability of finding any specific value of
$v_i$ does not depend on whether we are dealing with a peak or simply a point
with the same given (positive) value of $x$.

The correlations among variables $v_i$ take, for the Gaussian window, the
general form
\begin{equation}
\langle
v_i\,v_j\rangle=(-1)^{i+j+\delta_{0i}+\delta_{0j}}\,{\sigma_{i+j}^2\over
\sigma_{2i}\,\sigma_{2j}}
\end{equation}
($i,j\ge 0$), with $\delta_{ij}$ the Kr\"onecker delta. We must remark that,
although the notation used in the preceding equations presumes all variables
defined on scale $R$, they are also valid for variables defined on different
scales by just appropriately changing the values of the spectral moments
involved. In particular, were any variable, say $v_i$, defined on another
scale $R'$, one would be led to just the same expression for the correlations
as in equation (A11) but with $\sigma_i$ and $\sigma_{i+j}$ replaced by
$\sigma'_i\equiv \sigma_i(R')$ and $\sigma_{i+j\,\,h}\equiv
\sigma_{i+j}\{[0.5\,(R^2+{R'}^2)]^{1/2}\}$, respectively.

Thus, from equations (A8) to (A11) one can readily calculate the density
function ${\cal N}_{pk}$ of peaks at a fixed scale $R$ per infinitesimal
ranges of $\nu$, $x$, and any other set of the previous variables $v_i$
($i\ge 2$). And by integrating it over variable $\nu$ with the constraint
(see paper I)
\begin{equation}
{\delta\over \sigma_0}\le \nu \le
{\delta\over \sigma_0} + x\,{\sigma_2\over\sigma_0}\,R\,\Delta R,
\end{equation}
we can infer the density function $N_{pk}$ of peaks at a fixed $\delta$ per
infinitesimal ranges of $R$, $x$, and the same set of variables $v_i$ ($i\ge
2$). Consequently, such density functions $N_{pk}$ are just equal to
the corresponding ones ${\cal N}_{pk}$ times $x\,\sigma_2\,R/\sigma_0$.

Moreover, following the same scheme starting from the conditional
density analogous to the normal density in equation (A8) but for peaks
subject to any given constraint one is led to the same relation as above
between couples of conditional density functions ${\cal N}_{pk}$ and
$N_{pk}$ per infinitesimal ranges of any set of variables $v_i$ ($i\ge
2$) subject to that constraint.

\vspace{\baselineskip}
\noindent{\footnotesize A.3. NESTING PROBABILITIES}

Collapsing clouds associated with non-nested peaks with fixed $\delta$
yield a partition of space which makes the mass function of objects at $t$ be
correctly normalized (see paper I). This implies that the volume fraction
occupied by disjoint backgrounds with $\delta$ on filtering scales
between $R'$ and $R'+dR'$ or, equivalently, the probability to find a
point in any such backgrounds is $M(R')\,\rho^{-1}\,N(R',\delta)\,dR'$.
Therefore,
\begin{equation}
P(R',\delta|R,\delta)
\,dR'=\,{M(R')\over\rho}\,N(R',\delta)\,dR'\,
{N_{pk}(R,\delta|R',\delta)\over N_{pk}(R,\delta)}.
\end{equation}
gives the approximate probability that a typical (non-fissionable) peak with
$\delta$ on scale $R$ is nested within some non-nested peak with identical
density contrast but on a scale between $R'$ and $R'+dR'$ ($R<R'$), hereafter
simply called the (differential) nesting probability of a peak. (Remember
that the density of peaks that fission in the next $dR$ or $-d\delta$ is a
higher order differential.) This was used, in paper I, to derive equation
(1). Likewise, the nesting probability of peaks with given values of
variables $\delta$, $x$, and any set of variables $v_i$ ($i=2,...,n$ with
arbitrary $n$) is
\begin{equation}
P(R',\delta|R,\delta,x,v_2,...v_n)\,dR'
=\,{M(R')\over\rho}\,N(R',\delta)\,dR'\,
{N_{pk}(R,x,v_2,...,v_n,\delta|R',\delta)\over
N_{pk}(R,x,v_2,...,v_n,\delta)},
\end{equation}
with $N_{pk}(R,x,v_2,...,v_n,\delta|R',\delta)$ the conditional density
function analogous to $N_{pk}(R,\delta|R',\delta)$ in equation (A13) but per
infinitesimal ranges of the extra variables $x,v_2,v_3,...,v_n$. The last
factor on the right-hand member of equation (A14) satisfies the relation (see
Appendix A.2)
\begin{equation}
{N_{pk}(R,x,v_2,...,v_n,\delta|R',\delta)\over
N_{pk}(R,x,v_2,...,v_n,\delta)}={{\cal
P}(\nu',R'|\nu,x,v_2,...,v_n ,R)\, {\cal N}_{pk}
(\nu,x,v_2,...,v_n,R)\,x\,{\sigma_2\over\sigma_0}\,R\over {\cal
P}(\nu',R')\,{\cal N}_{pk}
(\nu,x,v_2,...,v_n,R)\,x\,{\sigma_2\over\sigma_0}\,R}
={{\cal P}(\nu',R'|\nu,x,v_2,...,v_n,R)\over {\cal P}(\nu',R')},
\end{equation}
with $\nu'\equiv \delta/\sigma_0'$, ${\cal P}(\nu',R'|\nu,x,v_2,...,v_n,R)
\,d\nu'$ defined in equation (A10), and ${\cal P}(\nu,R)\,d\nu$ the Gaussian
probability to find the scaled density contrast on scale $R$ between $\nu$
and $\nu+d\nu$. Notice that, for identical reasons as for the conditional
probability in equation (A10), the conditional probability given in equation
(A14) applies, in fact, to points. Note also that we are using the same
convention for the notation of the conditional probabilities as for the
density functions: they are denoted by a caligraphic capital p (in contrast
with the notation used in BBKS) when they are per infinitesimal range of
$\nu$ at a fixed $R$, and by a roman capital p when they are per
infinitesimal range of $R$ at a fixed $\delta$. Finally, it is worthwhile
mentioning that, contrary to what is suggested by the present notation, we
can write $\nu$ or $\delta$, indistinctly, when specifying the condition.

{}From equation (A10), it is clear that the nesting probability given by
equations (A14) and (A15) will depend on variables $v_i$ provided
only that these variables have non-null correlations with $\nu$ (or $\delta$)
and $x$ on scale $R$ or any explicit variable correlating with them. This is
what happens with the variables $v_i$ defined in equation (A7)
involving the different order scale derivatives of the density contrast.
(This is readily understood from the Taylor series expansion of $\nu$ on
scale $R'$ around $R$: the probability that a point has $\nu$ on scale $R'$
depends on the value of all scale derivatives of the density field on scale
$R$.) Thus, to accurately infer the density of nested peaks in any given peak
population one must calculate the distribution of these {\it infinite}
variables in that population. The only noticeable exception concerns the {\it
typical population\/} of peaks. The density of typical peaks at $\delta$ with
values of $v_i$ ($i=1,...,n$) in infinitesimal ranges,
$N_{pk}(R,x,v_2,...v_n,\delta) \,dR\,dx\,dv_2\,...dv_n\,$, times the nesting
probability $P(R',\delta|R,\delta,x,v_2,...v_n)\,dR'$ integrated over any
subset of variables $v_i$ coincides with the product of these two functions
without the explicit dependence on the integrated variables, as readily seen
from equation (A14) valid for any arbitrary values of the subindexes. In
particular, the integral over all $v_i$ for $i\ge 1$ is equal to the product
of $N_{pk}(R,\delta)\,dR$ times the reduced nesting probability
$P(R',\delta|R,\delta)\,dR'$ given in equation (A13), which justifies this
latter expression and equation (1).

The conditional probability ${\cal P}(\nu',R'|\nu,x,v_2,v_3,...,R )\,d\nu'$
extended to the infinite set of variables $v_i$ defined in equation (A7)
can be obtained according to equation (A10). After some lenghty algebra using
intermediate variables which only correlate with themselves (found by means
of the Gramm-Schmidt method) we arrive to the expression
\begin{equation}
{\cal P}(\nu',R'|\nu,x,v_2,v_3,...,R)\,d\nu'\,=\,{1\over
\sqrt{2\pi}\,\sigma_{\nu'}}
\,\exp \biggl[-{(\nu' -\bar\nu)^2\over
2\,\sigma^2_{\nu'}}\biggr]\,d\nu'
\end{equation}
with
\begin{displaymath}
\bar\nu\,=\,\alpha_{0} \nu + \alpha_{1} x +
\alpha_{2} v_2 + ...,
\end{displaymath}
\vspace{-\baselineskip}
\begin{equation}
\alpha_{i}=
\sum_{j=0}^\infty  \,B_j\,\beta_{i\,\,j-i}\,
\,\sum_{k=0}^j\,
\beta_{k\,\,j-k}\,<v'_0\,v_k>
\hskip 2truecm
\sigma_{\nu'}^2 =
1-\,\sum_{j=0}^\infty\,B_j
\,\biggl[\,\sum_{k=0}^j\,\beta_{k\,\,j-k}\,<v'_0\,v_k>
\biggr]^2
\end{equation}
($i\ge 0$), where a prime denotes the scale $R'$, the correlations in
angular brackets are given by equation (A11), and coeffecients $B_j$ and
$\beta_{k\,\,j-k}$ are defined as
\begin{equation}
(B_j)^{-1}=1+\sum_{k=0}^{j-1}\,\,\beta_{k\,\,j-k}\biggl(\beta_{k\,\,j-k}
+2\,\sum_{l=0}^{j-1-k}\,\beta_{j-l\,\,l}\,<v_k\,v_{j-l}>\biggr)\hskip 2cm
\beta_{i\,j-i}=-\sum_{k=0}^{j-1}
\,C_{i\,\,j-i}^k\,<v_j\,v_k>,
\end{equation}
($j\ge 1$ and $i\ne j$) in addition to $B_0=1$ and $\beta_{i\,0}=1$. In
equations (A18) we have used the notation
\begin{equation}
C_{i\,1}^k \equiv B_i\,\beta_{k\,\,i-k}\hskip2truecm
C_{i\,\,j-i}^k \equiv \sum_{l=k}^{j-1}\,
B_l\,\beta_{k\,\,l-k}\,\beta_{i\,\,l-i}
\end{equation}
for $i\le k$ and $C_{i\,j-i}^{k}\equiv C_{k\,j-k}^{i}$ for $i>k$.

{}From the general expressions of $\bar w$ and $\sigma_w^2$, equation (A10),
it can be shown, through the intermediate use of variables $\bar\nu^{(n)}$
defined as $\bar\nu$ but for just the first arbitrary $n+1$ variables $v_i$,
that
\begin{equation}
\langle \bar\nu\,v_i\rangle \,=\,\langle \nu'\,v_i\rangle \,\equiv\,
(-1)^{i+1+\delta_{0i}}\,{\sigma_{i\,h}^2\over \sigma'_{0}\,\sigma_{2i}},
\end{equation}
($i\ge 0$) which implies, on its turn, a similar relation for any (finite or
infinite) linear combination of $v_i$, in particular,
\begin{equation}
\langle \bar\nu^2\rangle \,=\,\langle
\nu'\,\bar\nu\rangle\,=\,1-\sigma_{\nu'}^2.
\end{equation}
On the other hand, taking into account that the correlation between two
variables is equal to the integral of the product of their Fourier transforms
taken at $\vec r=0$ and defining the new variable $\bar x$ as
$-(\sigma_2\,R)^{-1}\,\partial_R(\sigma'_0\,\bar\nu)$ we have
\begin{equation}
\langle \bar x\,v_i\rangle \,=\,{\sigma'_0\over\sigma_2}\,
\biggl({\sigma_{2i+1}^2\over\sigma_{2i}^2}-{1\over
R}\,\partial_R\biggr)\langle \bar\nu\,v_i\rangle
 + (-1)^{\delta_{i\,0}}\,{\sigma'_0\,\sigma_{2(i+1)}\over\sigma_2\,
\sigma_{2i}}\,\langle \bar\nu\,v_{i+1}\rangle
\end{equation}
($i\ge 0$) and
\begin{equation}
\langle \bar x\,\bar\nu\rangle \,=\,-\,{\sigma'_0\over\sigma_2}\,
{1\over 2\,R}\,\partial_R \langle \bar\nu^2\rangle.
\end{equation}
By substituting the correlations in equation (A20) into equation (A22) we
obtain
\begin{equation}
\langle\bar x\,v_i\rangle \,=\,0
\end{equation}
($i\ge 0$)
which also implies
\begin{equation}
\langle \bar x\,\bar\nu\rangle=0\hskip 2truecm\langle\bar x^2\rangle=0.
\end{equation}
Thus, $\langle\bar\nu^2\rangle$ and $\sigma_{\nu'}$ do not depend on $R$ (see
eqs. \lbrack A21\rbrack\ and \lbrack A23\rbrack) which ultimately implies
that $\sigma_{\nu'}$ is null and $\langle\bar\nu^2\rangle$ equal to unity.
Indeed, according to its definition, equation (A17), $\sigma_{\nu'}$ is null
for $R'=R$ and since it does not depend on $R$ it is necessarily null for any
value of $R'$. Then, equation (A16) leads to
\begin{equation}
{\cal P}(\nu',R'|\nu,x,v_2,v_3,...,R)\,d\nu'\,=\,\delta(\nu'-\bar\nu)\,d\nu'.
\end{equation}
This result is not surprising since fixing the values of the density contrast
and every order scale derivative of it on a given scale $R$ automatically
fixes, through the Taylor series expansion of $\delta$ as a function of the
filtering scale, the value of the density contrast in any other scale $R'$.
Finally, by substituting ${\cal P}$ given by equation (A26) into equation
(A15) and the latter into (A14) we arrive to the following expression for the
nesting probability
\begin{equation}
P(R',\delta|R,\delta,x,v_2,v_3,
...)\,dR'
\equiv P(R',\delta|\bar\nu)\,dR'=
\,{M(R')\over\rho}\,N(R',\delta)\,dR'\,
{\delta(\nu'-\bar\nu)\over {\cal P}(\nu')},
\end{equation}
with $\nu'=\delta/\sigma_0'$.

\vspace{\baselineskip}
\appendix \setcounter{section}{2}
\centerline{\footnotesize APPENDIX B: NET DENSITY OF PEAKS BECOMING NESTED}

The density of peaks at $\delta_f\equiv\delta-\Delta\delta$, with
$\Delta\delta$ positive and arbitrarily small, on scales between $R_f$ and
$R_f+dR_f$ and variables $x_f,{v_2}_f,{v_3}_f$,... in infinitesimal ranges,
which result by continuous evolution (hence, without fissioning) from peaks
at $\delta$ on scales between $R$ and $R+dR$ and $x,v_2,v_3,...$ in
infinitesimal ranges is
\begin{equation}
N_{pk}^{ev}(R_f,x_f,{v_2}_f,{v_3}_f,
...,\delta_f)\,dR_f\,dx_f\,d{v_2}_f\,...
=N_{pk}(R,x,v_2,v_3,...,\delta)\,dR\,dx\,dv_2\,dv_3\,...,
\end{equation}
with
\begin{equation}
R_f\approx R+{\Delta \delta\over x\,\sigma_2\,R}
\end{equation}
\begin{displaymath}
 x_f \approx
x+\biggl(v_2\,{\sigma_4\over\sigma_2}+x\,{\sigma_3^2\over\sigma_2^2}\biggr)
{\Delta \delta\over x\,\sigma_2} +\bigg(\sum_i{\partial_R \eta_i\over
\lambda_i}\,\partial_i x\biggr)\, {\Delta \delta\over x\,\sigma_2\,R}
\end{displaymath}
\begin{equation}
{v_2}_f \approx
v_2+\biggl[
v_3\,{\sigma_6\over\sigma_4}+v_2\biggl({\sigma_5^2\over
\sigma_4^2}-{1\over R^2}\biggr)\biggr] {\Delta \delta\over x\,\sigma_2}
+\bigg(\sum_i{\partial_R \eta_i\over
\lambda_i}\,\partial_i v_2\biggr)\, {\Delta \delta\over
x\,\sigma_2\,R}
\end{equation}
\vspace{-\baselineskip}
\begin{displaymath}
...\phantom{{\Delta \delta\over x\,\sigma_2}}
\end{displaymath}
to first order in $\Delta\delta$. Equation (B1) states that the density
$N_{pk}(R,x,v_2,v_3,...,\delta)$ $dR\,dx\,dv_2\,dv_3\,...$ of peaks is
conserved through continuous evolution from $\delta$ to $\delta_f$. Equations
(B2) and (B3) give the shift from $\delta$ to $\delta_f$ in the values of all
the relevant variables. It is important to outline that we need to know,
indeed, the values of all variables $R$ and $v_i$ ($i\ge 1$) at $\delta_f$ in
order to calculate the density of evolved peaks which are nested since, as
explained in Appendix A.3, the nesting probability depends explicitly on all
these variables differently distributed in $N_{pk}^{ev}$ than in $N_{pk}$ at
$\delta_f$. Equation (B2) arises from the derivative $dR/d\delta$ along
continuous peak trajectories, equal to $-(x\,\sigma_2\,R)^{-1}$ (see paper
I). While the shift in the variables $v_i$, equations (B3), is equal to the
sum of two terms: one coming from the scale derivative of each particular
variable, and a second one coming from the scalar product of its spatial
gradient times the shift in position of the new peak relative to the old one
(eq. \lbrack A4\rbrack). However, the nesting probability does not depend on
the variables $\partial_R \eta_i$ since these variables do not correlate with
$\nu,x,v_2,v_3,...$ (see Appendix A.3). Thus, in averaging below over all
variables, these second terms will contribute with a null mean (the
distribution of $\partial_R \eta_i$ for peaks is the same as for arbitrary
points). Consequently, we can drop these second terms which is equivalent to
taking an effective location for each evolved peak equal to the mean expected
value, that is, the same location as the original peak.

{}From equation (B1) we have that the density of peaks at $\delta$ per
infinitesimal ranges of $R$ and $x,v_2,v_3$,... which, after evolving to
$\delta_f$, are found to be nested (although not necessarily become
nested) into non-nested peaks with scales between $R'$ and $R'+dR'$ ($R\le
R_f< R'$) is
\begin{displaymath}
N_{pk}^{nest}(R\rightarrow
R',x,v_2,v_3,
...,\delta\rightarrow \delta_f)\,dR\,dR'\,dx\,dv_2\,dv_3\,...
\end{displaymath}
\vspace{-\baselineskip}
\begin{equation}
=\,N_{pk}(R,x,v_2,v_3,...,\delta)\,dR\,dx\,dv_2\,dv_3\,...\,
P(R',\delta_f|R_f,
\delta_f,x_f,{v_2}_f,{v_3}_f,...)\,dR',
\end{equation}
with $R_f$, $x_f$, ${v_2}_f$, ${v_3}_f$,... on the right hand side in terms
of $R$, $x$, $v_2$, $v_3$,... and $\Delta\delta$ through equations (B2) and
(B3), and the specific nesting probability $P$ given in Appendix A.3. To
obtain $N_{pk}^{nest}(R\rightarrow R',\delta\rightarrow \delta_f)\,dR\,dR'$
giving the density of peaks at $\delta$ per infinitesimal range of $R$ which
after evolving into $\delta_f$ are found to be nested into non-nested peaks
on scales between $R'$ and $R'+dR'$, we must integrate equation (B4) over
variables $x,v_2,v_3,...$. Given the simple expression of the nesting
probability appearing in equation (B4) in terms of the variable $\bar\nu$
(eq. \lbrack A27\rbrack) it is convenient to first transform $v_2$ and $v_3$
to $\bar\nu$ and $\bar x$. This can be done by repeated application of the
scheme given in Appendix A.2. Taking into account the correlations (A20) and
(A21) and the fact that $\sigma_{\nu'}^2=0$ (see the discussion after eq.
\lbrack A23\rbrack) we first obtain
\begin{equation}
N_{pk}(R,x,\bar \nu,R',\delta)\,dR\,dx\,d\bar\nu=N_{pk}(R,x,\delta)\,dR\,
dx\,\,{{\rm e}^{-\displaystyle{(\bar\nu-\bar\nu_*)^2\over
2\sigma_{\bar\nu}^2}}\over
\sqrt{2\pi}\,\sigma_{\bar\nu}}\,d\bar\nu,
\end{equation}
with $N_{pk}(R,x,\delta)={\cal N}_{pk}(\nu,x,R)\,x\,\sigma_2\,R/\sigma_0$ in
terms of the density function given in equation (A8) and
\begin{equation}
\bar\nu_*=\alpha_{\bar\nu\,0}\,{\delta\over\sigma_0}+
\alpha_{\bar\nu\,1}\,x\phantom{1^1_1\over1^1_1}\qquad
\sigma_{\bar\nu}^2=1-\epsilon^2\,\biggl[
{(1-\gamma^2\,r_1)^2\over 1-\gamma^2} + r_1^2\biggl]
\end{equation}
\vspace{-\baselineskip}
\begin{displaymath}
\alpha_{\bar\nu\,0}={\epsilon\,(1-\gamma^2
\,r_1)\over 1-\gamma^2}\qquad\qquad
\alpha_{\bar\nu\,1}=-{\epsilon\,\gamma\,(1-r_1)
\over 1-\gamma^2}
\end{displaymath}
with $\epsilon\equiv \sigma_{0\,h}^2/(\sigma_0\,\sigma_0')$ and $r_1\equiv
\sigma_{1\,h}^2\,\sigma_0^2/(\sigma_{0\,h}^2\,\sigma_1^2)$ already used in
BBKS. Then, from correlations (A24) and (A25) we are led to
\begin{equation}
N_{pk}(R,x,\bar \nu,\bar x,R',\delta)\,dR\,dx\,d\bar\nu\,d\bar
x= N_{pk}(R,x,\bar\nu,R',\delta)\,dR\,dx\,d\bar\nu\,\,\delta(\bar x)\,d\bar x.
\end{equation}
This result is well understood. As shown in Appendix A.3, variable $\bar\nu$
can only take the same value as $\nu'$ and, hence, $\sigma_0'\,\bar\nu$
can only depend on $R'$ so that $\bar x\equiv -(\sigma_2\,R)^{-1}\,\partial_R
(\sigma_0'\,\bar\nu)$ must be null. Finally, following the same procedure we
can infer the density of peaks for the remaining infinite series of variables
$v_4,v_5,...$ in infinitesimal ranges. But this is actually not necessary
since the integration of equation (B4) over these latter variables not
entering in $P$ is trivial, arriving to
\begin{equation}
N_{pk}^{nest}(R\hskip-3pt\rightarrow\hskip-3pt R',x,\bar \nu,\bar
x,\delta\hskip-3pt\rightarrow\hskip-3pt\delta_f)\,dRdR'dxd\bar\nu d\bar x
=N_{pk}(R,x,\bar
\nu,\bar x,R',\delta) \,P(R',\delta_f|\bar \nu_f)\,dRdR'dxd\bar\nu d\bar x,
\end{equation}
with $\bar\nu_f=\bar\nu -\bar x\,\Delta\delta/(x\,\sigma'_0)$ and
$P(R',\delta_f|\bar \nu_f)$ given, for variables with subindex $f$, by
equation (A27). Then, by integrating equation (B8) over $\bar x$ and
$\bar\nu$, which leads to
\begin{equation}
N_{pk}^{nest}(R\rightarrow R',x,\delta\rightarrow\delta_f)\,dR\,dR'\,dx
={M(R')\over \rho}\,N(R',\delta_f)\,N_{pk}(R,x,\delta|R',\delta_f) \,dR',
\end{equation}
taking the Taylor series expansions around $\delta$, keeping first order
terms in $\Delta \delta$, and integrating over $x$ (in the positive range)
we obtain (see eq. \lbrack A13\rbrack)
\begin{equation}
N_{pk}^{nest}(R\rightarrow R',\delta\rightarrow\delta_f)\,dR\,dR'
=N_{pk}(R,\delta) \,P(R',\delta|R,\delta)
\, \Bigl\{1 - \partial_{\delta_f} \ln
[N(R',\delta_f)\,N_{pk}(R,\delta|
R',\delta_f)]\Big|_{\delta_f=\delta}\,\Delta\delta\Bigr\}dR\,dR'.
\end{equation}
Equation (B10) can also be written, to first order in $\Delta\delta$, as
\begin{equation}
N_{pk}^{nest}(R\rightarrow R',\delta\rightarrow\delta_f)\,dR\,dR'
=N_{pk}(R,\delta)\,dR\,P(R',\delta_f|R,\delta)\,dR',
\end{equation}
with
\begin{equation}
P(R',\delta_f|R,\delta)\,dR'\,=\,
{M(R')\over\rho}\,N(R',\delta_f)\,dR'
\,{N_{pk}(R,\delta|R',\delta_f)\over N_{pk}(R, \delta)}
\end{equation}
giving the probability (only for $R'>R$; the case $R'=R$ being
excluded) that a (non-fissionable) peak with $\delta$ on scale $R$ is located
on a disjoint background with $\delta_f$ on scales between $R'$ and $R'+dR'$.
This is just what one would expect from the same arguments leading to the
nesting probability (A13). Notice however that, in contrast to that case, in
which $\delta_f=\delta$ so that the corresponding probability could only
include the nesting effect, the probability on the left hand side of equation
(B12) corresponding to $\delta_f\ne\delta$ will now include not only the
nesting of peaks having evolved from $R$ to $R_f<R'$ but also direct
evolution of peaks from $R$ to $R'$. While, by construction, the right-hand
member of equation (B10) only includes the former kind of effect. However, as
readily seen from a similar development as that leading to $N_{pk}^{nest}$,
the density of peaks with $R$ at $\delta$ which directly evolve into the
scale $R'$ at $\delta_f$ is of high order in $\Delta\delta/(R'-R)$.
Therefore, these two expressions coincide, for a given fixed difference
$R'-R$, to first order in $\Delta\delta$ as used in equation (B12). (This
reasoning shows, in particular, that eq. \lbrack B11\rbrack\ is not true in
general; it is only approximately satisfied for very small values of
$\Delta\delta$. The reason for this is clear. The spatial shift of evolving
peaks makes the different order scale derivatives of the density contrast of
the evolved peak at $\delta_f$ deviate from those of the initial point at the
same $\delta_f$ for arbitrarily large values of $\Delta\delta$ (see eq.
\lbrack B3\rbrack). The equality only holds to first order in $\Delta\delta$.
Therefore, the probability to find a peak with $R$ at $\delta$ located in a
disjoint background with $R'$ at $\delta_f$ is different, in general, from
the probability that the corresponding evolved peak at $\delta_f$ is located
in that background. These two probabilities only approximately coincide for
very small $\Delta\delta$.)

We are now ready to calculate the net density of peaks with $\delta$ on
scales between $R$ and $R+dR$ {\it becoming nested\/} into non-nested peaks
with $\delta_f\equiv \delta-\Delta\delta$ on scales between $R'$ and
$R'+dR'$, $N^d(R\rightarrow R',\delta\rightarrow\delta_f) \,dR\,dR'$.
(Superindex $d$ stands for destruction since these filtering events
correspond to true mergers, hence the destruction, of objects.) To do this we
must compute the density of peaks which, after evolving (without fissioning)
from $\delta$ to $\delta_f$, turn out to be nested into non-nested peaks with
such larger scales minus the density of (non-fissionable) peaks at $\delta$
which are nested in the ancestors of those disjoint backgounds at $\delta_f$.
(These ancestors concern, in principle, any kind of evolution,
although the contribution of direct evolution can be ignored in practice
since being of higher order in $\Delta\delta$). This correction can be
readily performed following the same procedure as leading to equation (1).
The result is the integral equation
\begin{equation}
N^d(R\rightarrow R',\delta \rightarrow \delta_f)=
N_{pk}^{nest}(R\rightarrow R',\delta\rightarrow\delta_f)
\int_R^{R'}
dR''\,{M(R'')\over \rho}\, N^d(R''\rightarrow R',\delta\rightarrow
\delta_f)\,N_{pk}(R,\delta|R'',\delta)
\end{equation}
for the unknown function $N^d(R\rightarrow R',\delta\rightarrow\delta_f )$ in
terms of $N_{pk}^{nest}(R\rightarrow R',\delta \rightarrow\delta_f)$ (eq.
\lbrack B10\rbrack) and $N_{pk}(R,\delta|R'',\delta)$ (see eq. \lbrack
1\rbrack).

But we do not need to solve equation (B13). Given the meaning of
$N^d$ and $N$, we have $N^d(R''\rightarrow R',\delta\rightarrow\delta)\equiv
N(R',\delta)\,\delta(R'-R'')$. (Notice that $R''$ reaches the value $R'$
inside the integral of eq. \lbrack B13\rbrack.) And since
$N_{pk}^{nest}(R\rightarrow R',\delta\rightarrow\delta)$ is equal to
$N_{pk}(R,\delta)\,P(R',\delta| R,\delta)$ (see eq. \lbrack B10\rbrack) we
are led to $N^d(R\rightarrow R',\delta\rightarrow \delta)\equiv 0$ for
$R<R'$ as in the present case. Thus, equation (B13) reduces to a simple
relation between first order terms in $\Delta\delta$. By dividing this
relation by $\Delta\delta$ we are led (see eqs. \lbrack B10\rbrack\ and
\lbrack A13\rbrack) to the Volterra type integral equation of the second kind
\begin{equation}
{\bf N}^d(R\rightarrow R',\delta) =
-\,{M(R')\over \rho}\,\partial_{\delta_f}
[N(R',\delta_f)\,N_{pk}(R,\delta|
R',\delta_f)]\Big|_{\delta_f=\delta}
- \int_R^{R'} dR''\,{M(R'')\over \rho}\, {\bf N}^d(R''\rightarrow
R',\delta)
\, N_{pk}(R,\delta|R'',\delta)
\end{equation}
whose solution gives the wanted net density ${\bf N}^d(R\rightarrow
R',\delta)\,dR\,dR'\,d\delta$ of peaks at $\delta$ with
scales between $R$ and $R+dR$ becoming nested (merging) into non-nested peaks
with scales between $R'$ and $R'+dR'$ in the next $-d\delta$. Equation (B14)
can be solved numerically by iteration from the initial approximate solution
$-M(R')\,\rho^{-1}\,\partial_{\delta_f} [N_{pk}(R',\delta)\,N(R,\delta|
R',\delta)]\big|_{\delta_f=\delta}$. Actually, this latter function is a very
good approximation to the wanted solution (with an error of less than 10\%
for the whole range of $R'$ in the least favorable case of $R$ close to the
lower limit of validity) for all power spectra analyzed. Thus, we can simply
take
\begin{equation}
{\bf N}^d(R\rightarrow R',\delta) \approx
-\,{M(R')\over \rho}\,\partial_{\delta_f}
[N(R',\delta_f)\,N_{pk}(R,\delta| R',\delta_f)]\Big|_{\delta_f=\delta}
\end{equation}
with $\partial_\delta N(R',\delta)$ given by numerical solution of the new
Volterra type integral equation
\begin{eqnarray}
\partial_\delta N(R',\delta) &=& \biggl[\partial_\delta
N_{pk}(R',\delta)-\int_{R'}^\infty dR''\,{M(R'')\over\rho}\,
N(R'',\delta)\,\partial_\delta N_{pk}(R',\delta|R'',\delta)\biggr]\nonumber
\\  &-& \int_{R'}^\infty dR''\,{M(R'')\over\rho}\,\partial_\delta
N(R'',\delta)\,N_{pk}(R',\delta|R'',\delta)
\end{eqnarray}
resulting from differentiation of equation (1).

\vspace{\baselineskip}
\appendix
\setcounter{section}{3} 
\centerline{\footnotesize APPENDIX C: NET DENSITY OF NON-NESTED
APPEARING PEAKS}

The net density of peaks appearing at $\delta$ with scales between
$R$ and $R+dR$ and variables $x$, $v_2$,$v_3$,... in infinitesimal ranges is
equal to the density of (non-fissionable) peaks with these characteristics
minus the density of peaks at $\delta$ with identical characteristics arising
by continuous evolution (hence, without fissioning) from peaks at
$\delta_i\equiv\delta+\Delta\delta$, with $\Delta\delta$ positive and
arbitrarily small. Therefore, the net density of {\it non-nested} peaks
appearing (forming) at $\delta$ with scales between $R$ and $R+dR$ is
\begin{displaymath}
{\bf N}^f(R,\delta,x,v_2,v_3,...)\,\Delta\delta
\,dR\,dx\,dv_2\,dv_3\,.... =\Bigl[1- P(R,\delta,x,v_2,v_3,...)\Bigr]
\end{displaymath}
\begin{equation}
\times \Bigl[N_{pk}(R,x,v_2,v_3,...,\delta)-N_{pk}^{ev}(R,x,v_2,v_3,
...,\delta)\Bigr]\,dR\,dx\,dv_2\,dv_3\,... ,
\end{equation}
with $P(R,\delta,x,v_2,v_3,...)$ the integral (over $R'>R$) of the
differential nesting probability given by equation (A14) and
$N_{pk}^{ev}(R,x,v_2,v_3,...,\delta)\,dR\,dx\,dv_2\,dv_3\,... $ the density
of evolved peaks, equal to
\begin{equation}
N_{pk}^{ev}(R,x,v_2,v_3,...,\delta)\,dR\,dx\,dv_2\,dv_3\,
...=N_{pk}(R_i, x_i,{v_2}_i,{v_3}_i,...,\delta_i)
\,dR_i\,dx_i\,d{v_2}_i\,d{v_3}_i...
\end{equation}
(Superindex $f$ stands for formation since these filtering events correspond
to the formation of new objects.) Equation (C2) is but equation (B1) for
the present notation, with the relations between variables with and without
subindex $i$ equal to the inverse of those given in equations (B2) and (B3).
Note that, for the same reasons as in Appendix B, we are forced to follow the
evolution of the whole infinite set of variables $x,v_2,v_3,...$.

{}From equations (C1) and (C2) we have
\begin{displaymath}
{\bf N}^f(R,\delta,x,v_2,v_3,...)
\,\Delta\delta\,dR\,dx\,dv_2\,dv_3\,
...=\Bigl[1- P(R,\delta,x,v_2,v_3,...)\Bigr]
\end{displaymath}
\begin{equation}
\times \Bigl[N_{pk}(R, x,v_2,v_3,...,\delta)- N_{pk}(R_i,
x_i,{v_2}_i,...,\delta_i)|J| \Bigr]\,dR\,dx\,dv_2\,dv_3\,...
\end{equation}
with $J$ the Jacobian of the transformation from variables with subindex $i$
to variables without subindex. Thus, by integrating equation (C3) over all
intermediate variables we obtain
\begin{displaymath}
{\bf N}^f(R,\delta) \,\Delta\delta\,dR= \biggl\{N(R,\delta) -
\int [1 - P(R,\delta,x,v_2,...)] \,N_{pk}(R_i,x_i,v_{2i}, ...,\delta_i)
\end{displaymath}
\begin{equation}
\times
\left[ 1 - {\Delta \delta \over
x_i R \sigma_2}\left(R{\sigma_3^2 \over \sigma_2^2} - {1 \over R}\right)
\right] \, dx_i\,dv_{2i}\,...\biggr\}\,dR.
\end{equation}
In deriving equation (C4) we have used the fact that $N_{pk}(R,\delta)$ times
the probability that a peak with $\delta$ on scale $R$ be non-nested, $1 -
P(R,\delta)$, is just equal to $N(R,\delta)$ (eqs. \lbrack A13\rbrack\ and
\lbrack 1\rbrack). We have also transformed the variables at $\delta$ in the
integral on the right hand side to variables at $\delta_i$, which balances
the Jacobian $|J|$, and then changed the differential $dR_i$ by $dR$
according to equation (B2).

For identical reasons as in Appendix B, it is convenient to express $N_{pk}$
and the nesting probability $P$ in equation (C4) in terms of variables
$\bar\nu$ and $\bar x$ instead of $v_2$ and $v_3$. After this substitution,
the integration over variables $v_4,v_5,...$ becomes trivial, and equation
(C4) can be written as
\begin{equation}
{\bf N}^f(R,\delta) \,\Delta\delta\,dR= \biggl\{N(R,\delta)
- \int [1 - P(R,\delta,\bar \nu)]\,  N_{pk}(R_i,x_i,\bar \nu_i,\bar
x_i,\delta_i )
\left[ 1 - {\Delta \delta \over x_i R \sigma_2}
\left(R{\sigma_3^2 \over \sigma_2^2} - {1
\over R}\right)\right]\, dx_i\,d \bar \nu_i\,d \bar x_i \biggr\}\,dR,
\end{equation}
where $R_i$ and $\delta_i$ are functions of $R$ and $x_i$ (see eq. [B2]) and
$\delta$, respectively. The integral involving only the density $N_{pk}$ is
straightforward, whereas that involving the product $P\,N_{pk}$ can be
calculated following the same steps as in Appendix B. By integrating over the
variables $\bar x_i$, $\bar \nu_i$ we are led to
\begin{equation}
{\bf N}^f(R,\delta) \,\Delta\delta\,dR= \biggl\{N(R,\delta)
- \int_0^\infty [1 - P(\delta|\delta_i,x_i,R_i)]\,  N_{pk}(R_i,x_i,
\delta_i)
\left[ 1 - {\Delta \delta \over x_i R \sigma_2}
\left(R{\sigma_3^2 \over \sigma_2^2} - {1
\over R}\right)\right]\, dx_i\biggr\}\,dR,
\end{equation}
with $P(\delta_f|\delta,x,R)\,dR$ the integral over $R'$ of the
probability $P(R',\delta|\delta_i,x_i,R_i)\,dR$ that a (non-fissionable) peak
with $R_i$ and $x_i$ at $\delta_i$ is located in a disjoint background with
$\delta$ on scales between $R'$ and $R'+dR'$ (eq. \lbrack B12\rbrack\ but
with the extra dependence on $x_i$ through the factor
$N_{pk}(R_i,\delta_i,x_i|R,\delta)/N_{pk}(R_i,\delta_i,x_i)$). Then taking
the Taylor series expansions around $\delta$, keeping first order terms in
$\Delta \delta$, integrating over $x_i$ (in the positive range), and
finally dividing by $-\Delta \delta$ and taking the limit $\Delta \delta
\rightarrow 0$ we arrive to the relation
\begin{displaymath}
{\bf N}^f(R,\delta)\, d \delta\,dR= \biggl\{
\partial_\delta N_{pk}(R,\delta) - \int_R^\infty \,dR'\,{M(R') \over
\rho}\,N(R',\delta)\,\partial_{\delta_i}N_{pk}(R,\delta_i| R',\delta)\Big|
_{\delta_i=\delta}
\end{displaymath}
\begin{equation}
- \partial_R {\cal N}_{pk}(\delta,R) +
\int_R^\infty \,dR'\,{M(R') \over \rho}\, N(R',\delta)\,
\partial_R {\cal N}_{pk}(\delta,R|\delta,R')
\biggr\}\,d \delta\,dR.
\end{equation}
{}For simplicity, we will take
\begin{equation}
{\bf N}^f(R,\delta)\, d \delta\,dR \approx
\left[\partial_\delta N_{pk}(R,\delta)- \partial_R
{\cal N}_{pk}(\delta,R)
\right]\,d \delta\,dR
\end{equation}
which is a reasonable approximation (with an error of less than 15 \% and 20
\% for scales within the range of validity of the model in the cases of the
CDM and the $n=-2$ power law spectra, respectively) to the exact relation
(C7). Note that, given the meaning of the densities $N_{pk}(R,\delta)\,dR$
and ${\cal N}_{pk}(\delta,R)\,d\delta$ (see paper I), expression (C8) is just
the density of peak trajectories leaving minus entering in the square
infinitesimal area of the $\delta$ vs. $R$ diagram, equal to the density of
peak trajectories appearing minus disappearing (i.e., fusioning minus
fissioning) in that inifitesimal area. That is, the approximation leading to
(C8) is but the neglect of the correction for nesting.

\vspace{\baselineskip}
\appendix
\setcounter{section}{4}
\centerline{\footnotesize APPENDIX D: MASS ACCRETION RATE}

{}From equation (B2) we have that the mass accreted from $\delta$ to
$\delta_f\equiv \delta-\Delta\delta$, with $\Delta\delta$ positive and
arbitrarily small, by any non-nested peak with initial
scale between $R$ and $R+dR$ and variable $x$ in an infinitesimal range is
\begin{equation}
\Delta M={dM\over dR}\,{1\over x\,\sigma_2\,R}\,\Delta\delta.
\end{equation}
The density of such accreting non-nested peaks is
\begin{equation}
N^a(R,x,\delta)\,dR\,dx=N(R,x,\delta)\,dR\,dx.
\end{equation}
In writing equation (D2) we have taken into account that the density of
non-nested peaks not accreting because merging from $\delta$ to
$\delta-\Delta\delta$ is a higher order correction (see the discussion
leading to eq. \lbrack B14\rbrack). Therefore, by dividing the density (D2)
by $N(R,\delta)\,dR$ we obtain the conditional probability
$p(x,R|\delta,R)\,dx$ that an accreting non-nested peak at $\delta$ with
scale $R$ has the appropriate value of $x$ in order to increase its mass by
$\Delta M$ given by equation (D1) in the passage from $\delta$ to
$\delta-\Delta\delta$. From equation (1) but for peaks with variable $x$ in
an infinitesimal range (or, equivalenly, from eq. \lbrack A14\rbrack, after
integrating the product $N_{pk}\,P$ over the remaining variables) this
conditional probability takes the form
\begin{equation}
p(x,R|\delta,R)
\,dx={dx\over
N(R,\delta)}\biggl[N_{pk}(R,x,\delta)-\int_R^\infty
dR'\,{M(R')\over\rho}\, N(R',\delta)\,N_{pk}(R,x,\delta|R',\delta)\biggr],
\end{equation}
with $N_{pk}(R,x,\delta)={\cal N}_{pk}(\delta,x,R)\,x\,\sigma_2\,R$ and
$N_{pk}(R,x,\delta|R',\delta)={\cal N}_{pk}(\delta,x,R|\delta,R')
\,x\,\sigma_2\,R$ in terms of the analogous density functions calculated by
BBKS (see eq. \lbrack A8\rbrack\ and paper I).

By changing variable $x$ into $M'=M+\Delta M$ we can compute the
instantaneous accretion rate of objects of mass $M$ per specific range of
mass $M'$ of the final object similar to the merger rate (6). By doing so we
arrive to the fact that this transition rate is identically null. The
reason for this is that, as mentioned in Appendix B, the density of peaks
with scales between $R$ and $R+dR$ at $\delta$ which directly evolve into
peaks with scales between $R'$ and $R'+dR'$ at $\delta_f$ is of higher order
than one in $\Delta\delta=\delta_f-\delta$. This result is well understood.
{}From the viewpoint of the accreting object, the process is not a transition
between two different masses but as a continuous mass increase. Consequently,
no discrete increment $\Delta M$ can be achieved in the limit
$\Delta\delta\rightarrow 0$. Of course, such a continuous evolution of the
accreting object during the small interval $\Delta t$ necessarily causes (is
made at the expense of) the merger (capture) of a number of tiny objects
which do make a finite transition in mass. And it is taking the limit for
vanishing $\Delta t$ of the change in the number density of these latter
objects that one obtains a non-vanishing accretion(+merger) rate (see \S\ 3).

In any event, there is no problem, even from the viewpoint of accreting
objects, in obtaining the instantaneous {\it mass accretion}
rate at $t$ for objects of mass $M$. Equation (D1) tells us that the
instantaneous mass increase rate, by accretion, for objects arising from
peaks with $\delta$ and the specific value of $x$ is
\begin{equation}
{dM\over dt}={dM\over dR}\,{1\over x\,\sigma_2\,R}\,\biggl|{d\delta_c\over
dt}\biggr|.
\end{equation}
Therefore, the instantaneous mass accretion rate at $t$ for objects of mass
$M$ (disregarding any other particularity) is the average of the specific
rate (D4) for the probability function (D3) with $R$ and $\delta=\delta_c$
expressed in terms of $M$ and $t$ through equations (3) and (4). By
performing this average we arrive to
\begin{equation}
r^a_{mass}(M,t)={dM\over dR}\,{1\over
x_{eff}\,\sigma_2\,R}\,\biggl|{d\delta_c\over dt}\biggr|,
\end{equation}
with
\begin{equation}
{1 \over x_{eff}} =
{1\over \langle x\rangle}\,\biggl[\,1+ \int_R^\infty
dR'\,{M(R')\over\rho}\, N(R',\delta)\,\biggl(1-{\langle x\rangle\over
\widetilde {\langle x\rangle}}\biggr)\,{N_{pk}(R,\delta|R',\delta)\over
N(R,\delta)}\biggr].
\end{equation}
Since $1/\langle x\rangle$ is a very good approximation to $1/x_{eff}$ (with
an error of less than 5\% for masses within the range of validity of the
model) we will simply take
\begin{equation}
r^a_{mass}(M,t) \approx {dM\over dR}\,{1\over
\langle x\rangle\,\sigma_2\,R}\,\biggl|{d\delta_c\over dt}\biggr|.
\end{equation}

\twocolumn
\def\subsection#1{\vspace{2mm}\parindent=32mm\footnotesize}

\normalsize
\end{document}